\documentclass[12pt]{iopart}
\usepackage{color}
\usepackage{graphicx}
\usepackage{amssymb}

\begin{document}

\newcommand{\ket}[1]{| #1 \rangle}
\newcommand{\bra}[1]{\langle #1 |}
\newcommand{\red}{\color[rgb]{0.8,0,0}}

\title[Finite-size atomic ensemble in a ring cavity]{Generating coherence and entanglement with a finite-size atomic ensemble in a ring cavity}

\author{Li-hui Sun$^{1,2}$, Gao-xiang Li$^{1}$, Wen-ju Gu$^{1}$ and Zbigniew Ficek$^{3}$}
\address{$^{1}$Department of Physics, Huazhong Normal University, Wuhan 430079, PR China}
\address{$^{2}$College of Physical Science and Technology, Yangtze University, Jingzhou, Hubei 434023, PR China}
\address{$^{3}$The National Centre for Mathematics and Physics, KACST, P.O. Box 6086, Riyadh 11442, Saudi Arabia}
\eads{\mailto{gaox@phy.ccnu.edu.cn}}

\begin{abstract}
We propose a model to study the coherence and entanglement resulting from the interaction of a finite-size atomic ensemble with degenerate counter-propagating field modes of a high-$Q$ ring cavity. Our approach applies to an arbitrary number of atoms $N$ and includes the spatial variation of the field throughout the ensemble. We report several new interesting aspects of coherence and entangled behaviour that emerge when the size of the atomic ensemble is not taken to the thermodynamic limit of $N\rightarrow\infty$. Under such conditions, it is found that the counter-propagating cavity modes, although in the thermodynamic limit are mutually incoherent and exhibit no one-photon interference, the modes can, however, be made mutually coherent and exhibit interference after interacting with a finite-size atomic ensemble. It is also found that the spatial redistribution of the atoms over a finite size results in nonorthogonality of the collective bosonic modes. This nonorthogonality leads to the super-bunching effect that the correlations of photons of the individual cavity modes and of different modes are stronger than those of a thermal field. We also investigate the spectral distributions of the logarithmic negativity and the variances of the output fields. These functions determine the entanglement properties of the output cavity fields and can be measured by a homodyne technique. We find that the entanglement is redistributed over several components of the spectrum and the finite-size effect is to concentrate the entanglement at the central component of the spectrum.
\end{abstract}

\pacs{42.50.Ar, 42.50.Pq, 42.70.Qs}

\submitto{\NJP}

\maketitle

\section{Introduction}\label{int}
Generation of continuous variable entangled states with atomic ensembles coupled to a radiation field has been intensively discussed both theoretically and experimentally in recent years~\cite{kp03,jz03,at03,pf04,bl05,fc05,wb10,mp08}. Atomic ensembles are macroscopic systems composed of a large number of atoms, and therefore it is a common practice in the theoretical treatments to work in the thermodynamic limit which takes the number of atoms $N$ inside an ensemble to infinity, $N\rightarrow\infty$. Under this approximation, the collective atomic operators are often represented, by using the Holstein-Primakoff representation of angular momentum operators~\cite{hp}, in terms of mutually independent bosonic modes, called collective bosonic modes. A large number of studies of a such system have been carried out in searching for superradiance and quantum phase transitions~\cite{de07,ac10,sc10}. The atomic ensembles have also been used to demonstrate the deterministic creation of nonclassical light fields in the interaction of atoms with a cavity field. Cavities, in particular microwave and ring cavities, provide efficient and controllable setting for a strong interaction between macroscopic atomic ensembles and the electromagnetic field~\cite{krb03,na03,kl06,gr00}. For example, Parkins {\it et al.}~\cite{ps06} have demonstrated that atomic ensembles interacting collectively with laser fields inside a high-$Q$ ring cavity can be unconditionally prepared in a two-mode squeezed state. The scheme, which is a generalization of the Guzman {\it et al.}~\cite{gr06} scheme to four-level atoms, is based on a suitable driving of the atomic ensembles with two external laser fields and coupling to a damped cavity mode that prepares the atoms in a pure squeezed (entangled) state. Similar schemes have been proposed to realize an effective Dicke model operating in the phase transition regime, to create a stationary subradiant state in an ultracold atomic gas~\cite{cb09}. This approach has also been considered as a practical scheme to prepare trapped and cooled ions in pure entangled vibrational states~\cite{li06} and to prepare four ensembles of hot atoms in pure entangled cluster states~\cite{lk09,gx10}. Recently, Krauter {\it et al.}~\cite{km10} have proposed to employ dissipation for generating a steady state entanglement between two distant atomic ensembles.

Studies of macroscopic systems composed of atomic ensembles interacting with a cavity field do not have to be confined to the thermodynamic limit. It has recently been demonstrated experimentally that small atomic ensembles could serve as a resource for quantum metrology and quantum information science~\cite{es08,gr10,re10,ls10}. This is the purpose of the present paper to consider a spatially extended finite-size atomic ensemble interacting with counter-propagating modes of a high-$Q$ ring cavity. Special emphasis is given to identifying intrinsically finite-size effects. The approach adopted here is based on the solution of the master equation of an effective two-level system involving ground states of the four-level atoms forming the atomic ensemble. The approach has similarities with some previous treatments, except that we introduce a spatial dependence of the interaction of the cavity modes and the laser fields with the atoms.

The spatial dependence arises naturally in the interaction of the fields with a finite-size atomic ensemble~\cite{eb70,vs10}, and the objective is to explore explicitly the issue of size effects in creation of coherence and entanglement in continuous variable systems. Examples of coherence processes are given to illustrate the effect of a finite size of the atomic ensemble  on creation of an entanglement between bosonic modes of the system. We find that the dynamics of the finite-size atomic ensemble differs qualitatively from those given in the thermodynamic limit. The inclusion of finite-size effects leads to a wide variety of unusual features. In particular, we find that collective bosonic modes of a finite-size atomic ensemble are not in general orthogonal to each other. In the course of the derivation of an effective Hamiltonian of the system, we observe that one of the finite-size effects is to create a direct coupling between the counter-propagating cavity modes. The mode nonorthogonality that couples the counter-propagating modes can drastically modify the property of the system.
The important modification is that the coupling lifts the degeneracy of the cavity modes and leads to significantly different statistical properties of the modes.  We present solutions for the second-order statistical moments of different modes of the system and find that the mode nonorthogonality gives rise to phase locking between the cavity counter-propagating modes, which leads to interesting first-order coherence effects. We also study the second-order correlation functions of the counter-propagating modes and show that the nonorthogonality leads to the super-bunching effect. In addition, we show that the nonorthogonality creates correlations that are necessary for entanglement between the intracavity modes. However, we find that the correlations created are not strong enough to produce and entangle between the cavity counter-propagating modes. We are therefore led to consider spectral distributions of the field variances and logarithmic negativity~\cite{ls09,lg11} and find that the two-mode squeezing and the  entanglement can actually be created between spectral components of the output cavity fields.

The paper is organized as follows. In section~\ref{sec2}, we describe in more detail the cavity and atomic ensemble under consideration. We derive an effective Hamiltonian of the system and show that the major finite-size effect is in the nonorthogonality of the collective bosonic modes. We then apply the Hamiltonian to derive the Heisenberg equations of motion for the field operators, and solve them in terms of the Fourier transform variables. Section~\ref{sec3} is devoted for the study of the mode nonorthogonality on coherence and entanglement properties of the counter-propagating cavity modes. In particular, in sections~\ref{sec3a} and~\ref{sec3b}, we analyze the first and second order coherence, respectively, between the counter-propagating cavity modes. We pay particular attention to the role of the mode nonorthogonality in the creation of coherence and correlations between the modes. Spectral distributions of the logarithmic negativity and the variances of the output fields are considered in section~\ref{sec3d}, where we illustrate the possibility of the creation of entanglement between spectral components of the output fields of the cavity modes. A summary of results is presented in section~\ref{sec4}. Finally, in the Appendix, we present analytical expressions for the steady-state mode occupation numbers, average amplitudes and correlations between the modes.

\section{Atomic system and Hamiltonian}\label{sec2}

\begin{figure}[thb]
\begin{center}
\begin{tabular}{c}
\includegraphics[height=7cm,width=0.65\columnwidth]{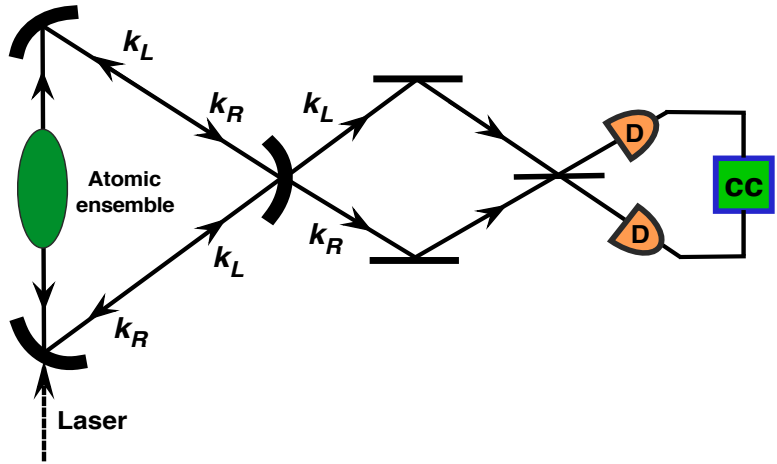}
\end{tabular}
\end{center}
\caption[nsa]{ \label{fig1a} A schematic diagram of a ring cavity containing an atomic ensemble trapped along the cavity axis. The clockwise and anti-clockwise cavity modes are damped with the same rate $\kappa$. The driving laser fields are injected through one of the cavity mirrors and co-propagate with the clockwise cavity mode. The output cavity modes are mixed at a $50/50$ beamsplitter and detected by two photodetectors. The output photocurrents are then registered by the coincidence counter CC.}
\end{figure}

The model we are considering is illustrated in Fig.~\ref{fig1a}. It consists of an atomic ensemble located inside a high-$Q$ ring cavity.  The cavity field is composed of two degenerate in frequency and overlapped counter-propagating modes, called clockwise $(R)$ and anti-clockwise~$(L)$ modes, characterized by equal frequencies $\omega_{R} = \omega_{L}\equiv \omega_{c}$, and anti-parallel wave vectors $\vec{k}_{R} = -\vec{k}_{L}\equiv \vec{k}_{c}$, respectively. The modes are represented by operators $\hat{a}_{R}\, (\hat{a}_{L})$ and  $\hat{a}_{R}^{\dag}\, (\hat{a}_{L}^{\dag})$ which are, respectively, the annihilation and creation operators for the cavity clockwise (anti-clockwise) mode.

\begin{figure}[thb]
\begin{center}
\begin{tabular}{c}
\includegraphics[height=8cm,width=0.55\columnwidth]{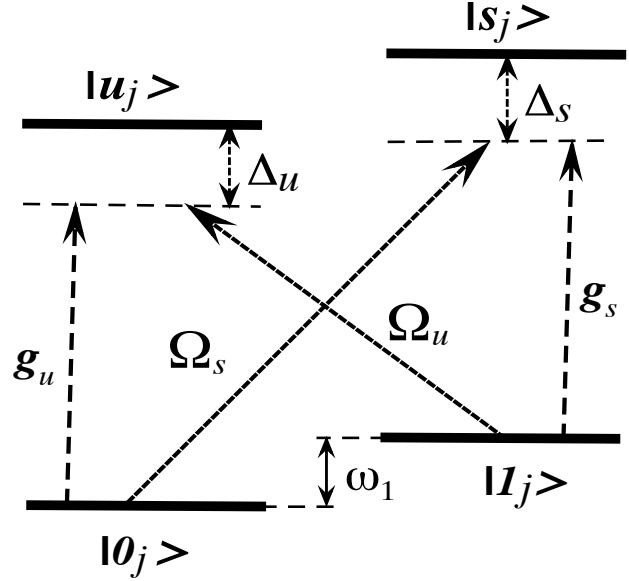}
\end{tabular}
\end{center}
\caption[nsa]{ \label{fig1b} Energy level scheme of the atoms and coupling configurations of the laser fields and the cavity modes. The laser fields of the Rabi frequencies $\Omega_{u}$ and $\Omega_{s}$ drive the atomic transitions $|1_{j}\rangle\rightarrow |u_{j}\rangle$ and $|0_{j}\rangle\rightarrow |s_{j}\rangle$, respectively. The atomic transitions~$|1_{j}\rangle\rightarrow |s_{j}\rangle$ and $|0_{j}\rangle\rightarrow |u_{j}\rangle$ are coupled to the cavity modes with the coupling strengths $g_{u}=g_{s}=g$.}
\end{figure}

The atomic ensemble is composed of $N$ identical four-level atoms interacting with external driving fields and a cavity field. An atom of the ensemble, say $j$th one, is represented by two non-degenerate ground states $\ket{0_{j}},\ket{1_{j}}$, two non-degenerate excited states $\ket{u_{j}}, \ket{s_{j}}$, and its position~$\vec{r}_{j}$, as illustrated in Fig.~\ref{fig1b}. In practice such a four-level system could correspond to an $F=1\leftrightarrow F^{\prime}=1$ transition as occurs in $^{87}$Rb atoms.

The cavity modes couple equally, i.e., with the same coupling strengths $g_{R}=g_{L}\equiv g$, to atomic transitions $\ket{0_{j}}\rightarrow\ket{u_{j}}$ and $\ket{1_{j}}\rightarrow \ket{s_{j}}$. This is acceptable since the degenerate overlapped cavity modes have the same polarization and geometry~\cite{krb03,na03,kl06}. In addition, the atomic ensemble is driven by pulse laser fields injected through one of the cavity mirrors and co-propagating with one of the cavity modes. The lasers are characterized by frequencies $\omega_{ls}$ and~$\omega_{lu}$, wave vectors $\vec{k}_{ls}=\vec{k}_{lu}\equiv \vec{k}_{l}$, and drive atomic transitions $\ket{0_{j}}\rightarrow\ket{s_{j}}$ and $\ket{1_{j}}\rightarrow \ket{u_{j}}$, with Rabi frequencies~$\Omega_{s}$ and~$\Omega_{u}$, respectively.

The total Hamiltonian for the atoms and the cavity modes can be written as
\begin{eqnarray}
\hat{H}_{T} = \hat{H}_{0}+\hat{H}_{AL}+\hat{H}_{AC} ,\label{hh1}
\end{eqnarray}
where
\begin{eqnarray}
\hat{H}_{0} &=&  \hbar\omega_c \left(\hat{a}_{R}^\dag \hat{a}_{R} +\hat{a}_{L}^{\dag} \hat{a}_{L}\right)
+\sum\limits_{j=1}^N \left(\hbar\omega_u|u_{j}\rangle\langle u_{j}|\right. \nonumber\\
&+&\left. \hbar\omega_s|s_{j}\rangle\langle s_{j}|
+\hbar\omega_1|1_{j}\rangle\langle 1_{j}|\right) \label{h01}
\end{eqnarray}
is the free Hamiltonian of the cavity modes and the atoms,
\begin{eqnarray}
\hat{H}_{AL} &=&  \frac{1}{2}\hbar \sum\limits_{j=1}^N \left\{\Omega_{u} {\rm e}^{i(\vec{k}_{l}\cdot\vec{r}_{j}
-\omega_{lu}t-\phi_{u})}|u_{j }\rangle\langle 1_{j }|\right. \nonumber\\
&+& \left. \Omega_{s}{\rm e}^{i(\vec{k}_{l}\cdot\vec{r}_{j}-\omega_{ls}t-\phi_{s})}|s_{j}\rangle\langle 0_{j}|
+{\rm H.c.}\right\} \label{hal1}
\end{eqnarray}
is the interaction Hamiltonian between the atoms and the driving fields, and
\begin{eqnarray}
\hat{H}_{AC} &=& \hbar g \sum\limits_{j=1}^N\left\{\left(\hat{a}_{R}{\rm e}^{i\vec{k}_{c}\cdot\vec{r}_{j}}
+ \hat{a}_{L}{\rm e}^{-i\vec{k}_{c}\cdot\vec{r}_{j}}\right)|u_{j }\rangle\langle0_{j }|\right. \nonumber\\
&+&\left. \left(\hat{a}_{R}{\rm e}^{i\vec{k}_{c}\cdot\vec{r}_{j}}
+ \hat{a}_{L}{\rm e}^{-i\vec{k}_{c}\cdot\vec{r}_{j}}\right)|s_{j }\rangle\langle1_{j }|+{\rm H.c.}\right\}
\label{hac1}
\end{eqnarray}
is the interaction Hamiltonian between the atoms and the two cavity modes. Here, $\phi_{u }$ and $\phi_{s}$ are phases of the laser fields, and $\omega_{1}, \omega_{s}$ and $\omega_{u}$ are atomic frequencies, corresponding to transitions $\ket{1_{j}}\leftrightarrow\ket{0_{j}}$,  $\ket{s_{j}}\leftrightarrow\ket{0_{j}}$, and $\ket{u_{j}}\leftrightarrow\ket{1_{j}}$, respectively. We have put zero energy at the ground state $\ket{0_{j}}$.

As we shall be interested in the generation of entanglement that requires minimal losses in the system, we consider an effective Hamiltonian in a dispersive regime that determines dynamics only between the ground states of the atoms. In this case, the cavity modes and the laser fields induce transitions between the ground states of the atoms via virtual transitions to far-off-resonant upper states. The effective Hamiltonian reads
\begin{eqnarray}
\hat{H}_{e} &= \hbar\omega \left(\hat{a}_{R}^\dag \hat{a}_{R} +\hat{a}_{L}^\dag \hat{a}_{L}\right)
+\hbar\, \alpha_{k}\delta \left(\hat{a}_{R}^{\dag}\hat{a}_{L} + \hat{a}_{L}^{\dag}\hat{a}_{R}\right) \nonumber\\
&+\hbar\omega_{0}\hat{J}_z +\left[\frac{\hbar\beta_u}{\sqrt{N}}\left(\hat{a}_{R}^\dag \hat{J}_{-k}
+\hat{a}_{L}^\dag \hat{J}_{+k}{\rm e}^{-i\phi_N}\right)+{\rm H.c.} \right]\nonumber\\
&+\left[\frac{\hbar\beta_s}{\sqrt{N}} \left(\hat{a}_{R}^\dag \hat{J}_{+k}^\dag {\rm e}^{i\phi_N}
+\hat{a}_{L}^\dag \hat{J}_{-k}^\dag\right)+{\rm H.c.} \right]  .\label{he2}
\end{eqnarray}
Detailed derivation of the effective Hamiltonian together with the definitions of the parameters involved is presented in Appendix A. Note that the advantage of working in the dispersive limit of large detunings $\Delta_{u}$ and $\Delta_{s}$ is to avoid spontaneous emission from the upper atomic states. In the derivation of (\ref{he2}), we have assumed further that the detunings $\Delta_{u}$ and $\Delta_{s}$ are much larger than the splitting of the ground states, i.e. $\Delta_{u},\Delta_{s}\gg \omega_{1}$. This allows us to ignore decoherence of the ground states due to elastic Rayleigh scattering. Recently, Uys {\it et al.}~\cite{ub10} have demonstrated, both theoretically and experimentally, that in the case of the detunings $\Delta_{u}$ and $\Delta_{s}$ comparable to the splitting of the ground states may result in a considerable Rayleigh decoherence in the system.

Among many parameters involved in the Hamiltonian (\ref{he2}), the most important for the purpose of the present paper is the parameter
\begin{eqnarray}
\alpha_{k}\delta =\alpha_{k} \frac{Ng^{2}}{\Delta}
\end{eqnarray}
which stands for the strength of the direct coupling between the cavity modes. The coupling is caused by the spatial variation of the cavity modes that arises from the interaction of the modes with the finite-size atomic ensemble. The spatial variation is completely determined by the parameter $\alpha_{k}$, which is of the from
\begin{eqnarray}
\alpha_{k} {\rm e}^{\pm i\phi_{N}} = \frac{1}{N}\sum\limits_{j=1}^{N} {\rm e}^{\pm 2i\vec{k}_{c}\cdot \vec{r}_{j}} .\label{alp}
\end{eqnarray}
This position dependent factor is recognized as the usual phase matching condition and represents an effective spread in phase difference between the cavity modes at $\vec{r}_{j}$. It follows that the factor will be different from zero when $N$ is not too large and $\vec{r}_{j}$ are small. It is easy to establish that the factor vanishes in the thermodynamic limit of $N\rightarrow\infty$.

The Hamiltonian (\ref{he2}) describes the interaction of a collection of $N$ two-level systems with the cavity counter-propagating modes. It involves linear interaction terms, proportional to $\beta_{u}$, as well as nonlinear interaction terms, proportional to $\beta_{s}$. Generally speaking, there are three different types of virtual transitions in the atoms; one is due to absorption of a photon of frequency $\omega_{ls}$ from a pulse laser accompanied by the emission of a photon to either $R$ or $L$ cavity mode. This process takes the atom from the state $\ket{0_{j}}$ to the state $\ket{1_{j}}$. The second process is due to absorption of a photon of frequency $\omega_{lu}$ from a pulse laser accompanied by the emission of a photon to either cavity mode $R$ or $L$. This process takes the atom from the state $\ket{1_{j}}$ to the state $\ket{0_{j}}$. Finally, the third process is due to absorption of a photon from either $R$ or~$L$ cavity mode accompanied by the emission of a photon of the same frequency to the counter-propagating mode. This process does not change the state of the atom.

The later process is the most interesting, because it is related to finite-size effects and is not encountered at all under the thermodynamic limit of $N\rightarrow\infty$. It shows that, after the interaction with the finite-size atomic ensemble, there is generally mutual coherence between the cavity modes. The parameter $\alpha_{k}\delta$ characterizes the strength of the coupling between the cavity modes and expresses the coherent exchange of photons between the modes. This simply reflects the presence of a phase relation between the counter-propagating cavity modes. The efficiency of the coupling depends on the parameter $\alpha_{k}$ which, according to~(\ref{alp}), is given by the phase mismatch of the propagation vectors of the cavity modes evaluated at the position of the individual atoms. The dependence of $\alpha_{k}$ on the phase mismatch factor $\vec{k}_{R}-\vec{k}_{L}=\pm 2\vec{k}_{c}$ indicates that for a given cavity mode, the other mode can be viewed as a 'phase-conjugate' field of the mode. The coupling happens because the counter-propagating cavity modes force an atom to move in the opposite directions. Since for a finite-size ensemble the force depends on the position of the atom, it creates a potential energy between atoms located at different positions. The energy averages to zero in the limit of $N\rightarrow \infty$ due to a random redistribution of the atoms inside the atomic ensemble.

Another interesting feature of a finite-size of the atomic ensemble is in the spatial dependence of the interaction between the atoms and the cavity fields that the multi-atom operators $\hat{J}_{\pm k}, \hat{J}^{ \dagger}_{\pm k}$ and $\hat{J}_{z}$ do not satisfy the standard angular momentum commutation relations. The reason is in the presence of the phase factors $\exp[i(\vec{k}_{l}\pm\vec{k}_{c})\cdot\vec{r}_{j}]$, which arise from the phase mismatch between the propagation direction of the cavity modes and directions of the laser fields. These factors represent an effective spread in phase difference between the laser and cavity fields at $\vec{r}_{j}$. As a consequence, the interaction is affected in a different way than the cavity modes. Moreover, the presence of two different phase mismatch factors indicates that the atomic ensemble may be coupled to the cavity modes in two distinctly different ways.

In order to explore this feature more explicitly, we adopt the Holstein-Primakoff representation of angular momentum operators~\cite{hp}, in which the two collective atomic operators,~$\hat{J}_{\pm k}$, are expressed in terms of annihilation operators~$\hat{C}_{\pm k}$ of the corresponding bosonic modes as follows:
\begin{eqnarray}
\hat{J}_{\pm k} = \sqrt{N} \hat{C}_{\pm k} ,\qquad \hat{J}_{z}
=\sum\limits_{j=1}^N \hat{b}_{j}^{\dag}\hat{b}_{j} ,\label{jmkz2}
\end{eqnarray}
where
\begin{eqnarray}
\hat{C}_{\pm k}=\frac{1}{\sqrt{N}}\sum\limits_{j=1}^N \hat{b}_{j}{\rm e}^{i(\vec{k}_{l}\pm\vec{k}_{c})\cdot\vec{r}_{j}} ,\label{cmk}
\end{eqnarray}
are collective bosonic operators with the annihilation $\hat{b}_{j}$ and creation~$\hat{b}^{\dagger}_{j}$ operators obeying the standard bosonic commutation relation $[\hat{b}_{j}, \hat{b}^{ \dagger}_{\ell}]=\delta_{j\ell}$.
It is easily verified that the collective bosonic operators do not in general commute, i.e.
\begin{eqnarray}
\left[\hat{C}_{\pm k},\hat{C}^{\dag}_{\mp k}\right] = \alpha_{k} {\rm e}^{\pm i\phi_N} .\label{commutec}
\end{eqnarray}
Again, the reason is in the presence of the position dependent phase factors. Hence, the collective modes of a finite-size atomic ensemble are not orthogonal to each other. The degree of nonorthogonality of the modes is determined by the phase matching parameter $\alpha_{k}$, and the modes become orthogonal in the thermodynamic limit of $N\rightarrow\infty$.

The commutation relation can also be viewed as a non-distinguishability criterion for the collective modes. In the thermodynamic limit, $\alpha_{k}=0$, and then the modes are completely distinguishable. For $\alpha_{k}\neq 0$, the modes are partly distinguishable and become completely indistinguishable when $\alpha_{k}=1$. As we shall see below, the non-orthogonality and thus indistinguishability of the modes will result in correlations between different modes of the system.

Before proceeding further, we note here that the bosonic representation of the collective atomic operators of a finite-size atomic ensemble places no restriction on the number of atoms composing the ensemble~\cite{pc08}. The representation is valid for an arbitrarily small number of atoms with the condition of a very low excitation probability of each atom, i.e. $\langle \sigma_{11}^{j}\rangle\ll 1$, where $\sigma_{11}^{j} =\ket{1_{j}}\bra{1_{j}}$.

The effective Hamiltonian (\ref{he2}) expressed in terms of the collective bosonic operators describes the interaction of a "fictitious" bosonic system with the two-mode cavity field, and has the form
\begin{eqnarray}
\hat{H}_{e} &= \hbar\omega \left(\hat{a}_{R}^\dag \hat{a}_{R} +\hat{a}_{L}^\dag \hat{a}_{L}\right)
+\hbar\, \alpha_{k}\delta \left(\hat{a}_{R}^{\dag}\hat{a}_{L} + \hat{a}_{L}^{\dag}\hat{a}_{R}\right) \nonumber\\
&+ \hbar\omega_{0}\hat{J}_z +\left[\hbar\beta_{u}\left(\hat{a}_{R}^{\dag}\hat{C}_{- k}
+\hat{a}_{L}^{\dag} \hat{C}_{+ k}{\rm e}^{-i\phi_N}\right)+{\rm H.c.} \right]\nonumber\\
&+\left[\hbar\beta_{s}\left(\hat{a}_{R}^\dag \hat{C}_{+ k}^\dag {\rm e}^{i\phi_N}
+\hat{a}_{L}^\dag \hat{C}_{- k}^{\dag}\right)+{\rm H.c.} \right] .\label{he2a}
\end{eqnarray}
Note that the Hamiltonian is symmetric under reversal of the direction of propagation of either the laser field or the cavity mode.

Instead of working with the collective operators $\hat{C}_{\pm k}$, we shall find convenient to work with two operators
\begin{eqnarray}
\hat{d}_{1} &= \frac{1}{\sqrt{2(1+\alpha_{k})}}\left(\hat{C}_{-k} + {\rm e}^{-i\phi_N}\hat{C}_{+k}\right) ,\nonumber\\
\hat{d}_{2} &= \frac{1}{\sqrt{2(1-\alpha_{k})}}\left(\hat{C}_{-k} - {\rm e}^{-i\phi_N}\hat{C}_{+k}\right) ,\label{d12}
\end{eqnarray}
which are linear symmetric and antisymmetric superpositions of the bosonic collective operators, respectively.
It is easily checked that the superposition operators are orthogonal to each other and obey the standard bosonic commutation relations, $[\hat{d}_i,\hat{d}_j]=0$ and $[\hat{d}_i,\hat{d}_j^\dag]=\delta_{ij}$. In terms of the superposition operators~(\ref{d12}), the effective Hamiltonian (\ref{he2a}) simplifies~to
\begin{eqnarray}
\hat{H}_{e} &= \hbar \left(\omega +\alpha_{k}\delta\right) \hat{a}_1^\dag \hat{a}_1 +\hbar\left(\omega -\alpha_{k}\delta\right) \hat{a}_2^\dag \hat{a}_{2} \nonumber\\
&+\hbar\omega_0\left( \hat{d}_1^\dag \hat{d}_1+\hat{d}_2^\dag \hat{d}_{2} \right)
 +2\hbar\lambda_{1}\, \hat{a}_{1x}\hat{d}_{1x} +2\hbar\lambda_{2}\, \hat{a}_{2y}\hat{d}_{2y} ,\label{he3}
\end{eqnarray}
where $\lambda_{1} =\beta\sqrt{1+\alpha_{k}}$, $\lambda_{2} =\beta\sqrt{1-\alpha_{k}},\, (\beta =\beta_s=\beta_u)$, and
\begin{eqnarray}
\fl \hat{a}_{1x} = \frac{\left(\hat{a}_1+\hat{a}_1^\dag \right)}{\sqrt{2}} ,\quad \hat{a}_{2y} = \frac{i\left(\hat{a}_{2}^\dag - \hat{a}_{2} \right)}{\sqrt{2}} ,\quad
\hat{d}_{1x} = \frac{\left(\hat{d}_1+\hat{d}_1^\dag \right)}{\sqrt{2}} ,\quad \hat{d}_{2y} = \frac{i\left(\hat{d}_{2}^\dag - \hat{d}_{2} \right)}{\sqrt{2}} ,
\end{eqnarray}
are in-phase and out-of-phase quadrature components of the cavity modes and the bosonic field operators, with
\begin{equation}
\hat{a}_1=\frac{(\hat{a}_{R}+\hat{a}_{L})}{\sqrt{2}} ,\qquad \hat{a}_2 =\frac{(\hat{a}_{R} -\hat{a}_{L})}{\sqrt{2}} .\label{a12}
\end{equation}

It is seen from~(\ref{he3}) that one of the finite-size effects on the system is to lift the degeneracy of the cavity modes by creating linear symmetric and antisymmetric superpositions of the modes with frequencies $\omega +\alpha_{k}\delta$ and $\omega -\alpha_{k}\delta$, respectively. It is also seen that the superposition collective modes are degenerate in frequency, but they do not behave similarly. The modes couple to the cavity superposition modes with different coupling strengths. The symmetric mode $\hat{d}_{1}$ couples to the cavity mode~$\hat{a}_{1}$ with an enhanced coupling strength $\lambda_{1}=\beta\sqrt{1+\alpha_{k}}$, whereas the antisymmetric mode $\hat{d}_{2}$ couples to the mode $\hat{a}_{2}$ with a reduced strength $\lambda_{2}=\beta\sqrt{1-\alpha_{k}}$. It is interesting to note that the pairs of modes $(\hat{a}_{1},\hat{d}_{1})$ and $(\hat{a}_{2},\hat{d}_{2})$ are decoupled from each other. This means that each pair can be independently prepared in a desired state. While the cavity superposition modes result from the linear coupling between the cavity counter-propagating modes, the collective bosonic modes couple to the cavity modes in linear as well as in a nonlinear way. This is the nonlinear coupling that may create entanglement between the cavity and the collective bosonic modes.

In the physical terms, the Hamiltonian (\ref{he3}) contains terms describing four-wave mixing of the up-shifted (signal) and down-shifted (idler) cavity modes with the degenerate collective modes. Other terms proportional to the products of creation and annihilation operators for the same mode result in a dispersive effect. There are also terms that couple creation and annihilation operators of the cavity modes with the creation and annihilation operators of the collective modes. This interaction is responsible for the back-action evading nature of quantum non-demolition detection.

It is worthwhile noting that, in spite of the fact that the finite-size feature of the system is manifested by the presence of three phase mismatch factors, the difference between the Hamiltonians of finite- and infinite-size atomic ensembles is embodied in a single parameter~$\alpha_{k}$. In other words, the dynamics of the system are independent of the direction of propagation of the laser fields. They depend solely on the phase mismatch of the cavity counter-propagating modes. It is only the presence of~$\alpha_{k}$ that pulls of the degenerate cavity modes above and below their resonance by equal amounts, $\delta_{c}$, and introduces an asymmetry to the coupling constants of the collective bosonic modes to the cavity modes.

Before moving on to the consideration of coherence and correlation features in the system, we first briefly comment about the threshold behaviour of the Hamiltonian (\ref{he3}). One can notice that the Hamiltonian (\ref{he3}) is of the form of two independent one-mode Dicke models
\begin{eqnarray}
\hat{H}_{e} = \hat{H}_{1} + \hat{H}_{2} ,\label{he3aa}
\end{eqnarray}
where
\begin{eqnarray}
\hat{H}_{1} &= \hbar\Omega_1\hat{a}_1^\dag \hat{a}_1 +\hbar\omega_0 \hat{d}_1^\dag \hat{d}_1
 +2\hbar\lambda_{1}\, \hat{a}_{1x}\hat{d}_{1x} ,\nonumber\\
\hat{H}_{2} &= \hbar\Omega_2\hat{a}_2^\dag \hat{a}_2 +\hbar\omega_0 \hat{d}_2^\dag \hat{d}_2
 +2\hbar\lambda_{2}\, \hat{a}_{2y}\hat{d}_{2y} .
\end{eqnarray}
with $\Omega_1=\omega +\alpha_{k}\delta$ and $\Omega_2=\omega -\alpha_{k}\delta$. Hence, many features predicted previously by other authors for the one-mode Dicke model in the thermodynamic limit can also be seen in our model~\cite{de07,ps06,gr06}. However, instead of focusing on these one-mode features, we prefer to specialize our considerations to novel features of the two-mode Dicke model that might be brought by finite-size effects. For example, there might be coherence and correlations existing between modes that are simultaneously involved in both Dicke models. We examine these properties shortly, but first we examine a manifestation of the finite-size effects in the threshold behaviour of the system. It is easy to see that in the case of $\alpha_{k}\neq 0$, the coupling strength $\lambda_{1}\neq \lambda_{2}$. As a consequence, there are two rather than one critical values of~$\beta$:
\begin{eqnarray}
\beta_{c1} &= \frac{1}{2\sqrt{1+\alpha_{k}}}\sqrt{\frac{\omega_0}
{\Omega_1}\left(\kappa^2+\Omega_1^2\right)} ,\nonumber\\
\beta_{c2} &= \frac{1}{2\sqrt{1-\alpha_{k}}}\sqrt{\frac{\omega_0}
{\Omega_2}\left(\kappa^2+\Omega_2^2\right)} .\label{betac12}
\end{eqnarray}
Thus, an interesting notable feature of the finite-size effects is the existence of two distinctive critical values of the coupling strength $\beta$. It is easily verified that the critical values~$\beta_{c1}$ and~$\beta_{c2}$ shift in opposite directions as $\alpha_{k}$ increases. Note that in the limit of $\alpha_{k}\rightarrow 1$, $\beta_{c1}$ approaches a finite value, whereas~$\beta_{c2}$ goes to infinity.

The existence of the two threshold values for $\beta$ indicates that the properties of the system could be different for different values of $\beta$. It what follows, we confine our considerations to the case of below the thresholds, i.e. $\beta <\beta_{c1}$.

\section{Coherence and entanglement induced by the finite-size effects}\label{sec3}

We now proceed to discuss the coherence and correlation features of the cavity modes brought by the finite-size effects of the atomic ensemble. As we have already mentioned, coherence and correlations can be created between different modes of the system. Here, we confine ourselves to the study of the coherence and correlations of the cavity counter-propagating modes only. The reason is that properties of the cavity modes can be directly measured by detecting of the out-put cavity fields. The coherence properties of the other modes of the system can be found from the properties of the out-put cavity fields.

In order to keep the considerations close to practical situations, we include a possible loss of cavity photons due to the damping of the cavity mode. With the cavity damping included, the state of the system is a statistical mixture determined by the density operator $\rho$ whose time evolution is governed by the master equation
\begin{eqnarray}
\dot{\rho} = -\frac{i}{\hbar}[\hat{H}_{e},\rho] + \frac{1}{2}\kappa\sum\limits_{j=1}^{2} \left(2\hat{a}_{j}\rho \hat{a}^\dag_{j}-\hat{a}^\dag_{j}\hat{a}_{j}\rho -\rho \hat{a}^\dag_{j} \hat{a}_{j} \right) ,\label{rhoc}
\end{eqnarray}
where $\hat{H}_{e}$ is given in (\ref{he3}) and $\kappa$ is the cavity damping rate. This is the only damping we consider here as we have already eliminated spontaneous emission from the atoms by choosing large detunings of the driving lasers and the cavity modes.

Our treatment is based on the solution of the Heisenberg equations of motion for the mode operators that are readily obtained from the master equation~(\ref{rhoc}), and are given by
\begin{eqnarray}
\dot{\hat{a}}_j(t)&=&-i[\hat{a}_j(t),\hat{H}_e]
-\kappa \hat{a}_j(t)+\sqrt{2\kappa}\,\hat{a}_{j}^{in}(t) ,\nonumber\\
\dot{\hat{d}}_j(t)&=&-i[\hat{d}_j(t),\hat{H}_e] ,\qquad  j =1,2,\label{lambqm}
\end{eqnarray}
along with the corresponding equations for the adjoint operators. In these equations, the operator $\hat{a}_{j}^{in}(t)$ describes the quantum noise injected at the cavity input.

It is easy to show that the set of differential equations for the mode operators splits into two independent sets, each composed of four coupled differential equations. The sets of the differential equations are conveniently solved by taking the Fourier transform of the operators
\begin{eqnarray}
\label{Fourier}
\tilde{u}(\nu)&=&\frac{1}{\sqrt{2\pi}}\int^\infty_{-\infty}{\rm e}^{i\nu t}\hat{u}(t)dt, \\
\tilde{u}^\dagger(-\nu)&=&\frac{1}{\sqrt{2\pi}}\int^\infty_{-\infty}{\rm e}^{i\nu t} \hat{u}^\dagger(t)dt,
\end{eqnarray}
where $\hat{u}$ denotes any one of the operators $\hat{a}_1,\hat{a}_2,\hat{d}_1,\hat{d}_2$ and $\hat{a}_{1}^{in},\hat{a}_{2}^{in}$. Expressed as equations for the transforms of the operators, the solution for the mode operators may be written~as
\begin{eqnarray}
\label{solq}
\tilde{d}_1(\nu)&=&\frac{\lambda_{1} \left[\tilde{a}_1(\nu) +\tilde{a}^\dag_1(-\nu)\right]}{(\nu -\omega_0)} ,\nonumber\\
\tilde{d}_2(\nu)&=&\frac{\lambda_{2} \left[\tilde{a}_2(\nu) -\tilde{a}^\dag_2(-\nu)\right]}{(\nu -\omega_0)} ,\nonumber\\
\tilde{a}_1(\nu)&=&\frac{M_{11}(\nu) \tilde{a}_{1}^{in}(\nu) + M_{12}(\nu) \tilde{a}^{in \dagger}_{1}(-\nu)}{D_1(\nu)} ,\nonumber\\
\tilde{a}_2(\nu)&=&\frac{M_{21}(\nu) \tilde{a}_{2}^{in} (\nu) - M_{22}(\nu) \tilde{a}^{in \dagger}_{2}(-\nu)}{D_2(\nu)} ,
\end{eqnarray}
where
\begin{eqnarray}
D_j(\nu) &=[\kappa - i(\nu - \Omega_j)][\kappa - i(\nu+\Omega_j)] (\nu^2 - \omega^2_0)
 + 4\lambda^2_j \omega_0 \Omega_j ,\nonumber\\
M_{j1}(\nu)&= \sqrt{2\kappa}\left\{[\kappa-i(\nu+\Omega_j)](\nu^2-\omega^2_0)
-2i\lambda^2_{j} \omega_{0}\right\} ,\nonumber\\
M_{j2}(\nu)&= -2i\sqrt{2\kappa}\,\lambda^2_j\omega_0 ,\qquad j=1,2 .\label{lamda}
\end{eqnarray}
The frequency dependent solution~(\ref{solq}) will be used for the calculations of the collective modes and the field correlation functions necessary for evaluation of the two-mode squeezing and entanglement spectra.

Since we are also interested in the steady-state coherence between the modes, we transform the solutions~(\ref{solq}) back to the time domain and take the steady-state limit. Then all the functions necessary for the explicit calculation of the coherence are obtained by the average over the initial vacuum state with zero occupation numbers for all the modes of the system. The steady-state solution for the field averages and correlation functions are listed in the Appendix~B.

\subsection{First-order coherence of the cavity modes}\label{sec3a}

First, we consider the first-order mutual coherence between the counter-propagating cavity modes. The mutual coherence between the cavity modes $\hat{a}_{R}$ and $\hat{a}_{L}$ is measured by the cross correlation $\langle \hat{a}_{R}^{\dag}\hat{a}_{L}\rangle$, the so-called coherence function, where the average is taken over the initial vacuum state of the modes~\cite{mw95}. The degree of coherence between the modes is given by
\begin{eqnarray}
\left|\gamma_{(R,L)}\right| = \frac{|\langle \hat{a}_{R}^{\dag}\hat{a}_{L}\rangle|}{\langle \hat{a}_{R}^{\dag}\hat{a}_{R}\rangle^{1/2}\langle \hat{a}_{L}^{\dag}\hat{a}_{L}\rangle^{1/2}} .\label{eq32}
\end{eqnarray}
The visibility ${\cal V}$ of the interference pattern, on the other hand, is given by
\begin{eqnarray}
{\cal V}_{(R,L)} = \frac{2|\langle \hat{a}_{R}^{\dag}\hat{a}_{L}\rangle|}{\langle \hat{a}_{R}^{\dag}\hat{a}_{R}\rangle + \langle \hat{a}_{L}^{\dag}\hat{a}_{L}\rangle} .
\end{eqnarray}
The degree of coherence and the visibility of the stationary cavity fields can be readily calculated using the steady-state solutions~(\ref{B1}). Since
\begin{eqnarray}
\langle \hat{a}_{R}^{\dag}\hat{a}_{R}\rangle = \langle \hat{a}_{L}^{\dag}\hat{a}_{L}\rangle =\frac{1}{2}\left(\langle \hat{a}_{1}^{\dag}\hat{a}_{1}\rangle + \langle \hat{a}_{2}^{\dag}\hat{a}_{2}\rangle\right) ,\label{e32a}
\end{eqnarray}
we see that the visibility equals to the degree of coherence independent of the parameters of the system. Moreover,
\begin{eqnarray}
\langle \hat{a}_{R}^{\dag}\hat{a}_{L}\rangle =\frac{1}{2}\left(\langle \hat{a}_{1}^{\dag}\hat{a}_{1}\rangle - \langle \hat{a}_{2}^{\dag}\hat{a}_{2}\rangle\right) = \alpha_{k}U(\omega,\kappa) ,\label{arl}
\end{eqnarray}
where
\begin{eqnarray}
U(\omega,\kappa) = \frac{w_1 (\alpha_{k}^{2}w_2 w_3 +u_2 u_3) + u_1 (u_3 w_2 + u_2 w_3 )}{(\omega^2-\alpha_{k}^2 \delta^2) (u_4^2-\alpha_{k}^2 w_4^2)} ,\label{e33a}
\end{eqnarray}
with
\begin{eqnarray}
u_{1} &= \beta^2 \left(\omega -\alpha_{k}^2 \delta \right),\qquad u_{2}=\alpha_{k}^2 \delta ^2+\kappa^2+\omega^2 ,\nonumber\\
u_3 &= u_2 \omega_0 - 4\beta^{2} \left(\alpha_{k}^2 \delta +\omega\right) ,\nonumber\\
u_4&=\alpha_{k}^{2}\delta \left(\omega_{0}\delta - 4\beta^2\right) - 4\omega_{0} \left(\beta^2  +\kappa^2 +\omega_{0}^2\right) ,\nonumber\\
w_1 &=\beta ^2 (\omega -\delta ) ,\qquad w_2=2\omega\delta ,\qquad
 w_3 =4 \beta^2 (\omega +\delta) - w_2 \omega_0 ,\nonumber\\
w_4&=4 \beta^2\left(\omega_{0}+\delta\right) -2 \delta \omega_{0}^2 .\label{e34a}
\end{eqnarray}
We see that the coherence function depends directly on the finite-size parameter $\alpha_{k}$. This implies that the cavity modes are correlated only when $\alpha_{k}\neq 0$. The mode nonorthogonality can transfer photons from one mode to the other. Thus, one of the aspects of finite-size effects is the creation of the first-order correlation between the cavity counter-propagating modes. This feature is not encountered at all under the thermodynamic limit of $N\rightarrow\infty$. In physical terms, we may attribute the appearance of the coherence to the fact that the counter-propagating modes are unresolved at the atomic ensemble, that it is impossible to tell to which mode the photon was emitted.
\begin{figure}[hbt]
\begin{center}
\begin{tabular}{c}
\includegraphics[height=8cm,width=0.75\columnwidth]{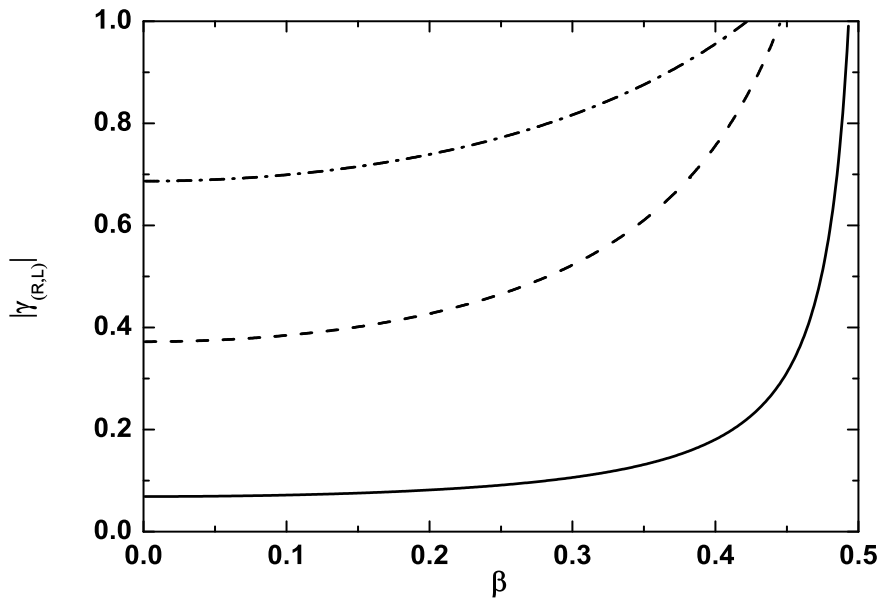}
\end{tabular}
\end{center}
\caption[nsa]{The steady-state degree of coherence $|\gamma_{(R,L)}|$ plotted as a function of the coupling strength $\beta$ for $\omega_0=\omega=1,\delta=0.1 \pi,\kappa=0.2$, and different degrees of mode nonorthogonality $\alpha_{k}$: $\alpha_{k}=0.1$ (solid line), $\alpha_{k} =0.5$ (dashed line), and $\alpha_{k} =0.8$ (dashed-dotted line). In this and in all the following figures, the parameters are normalised to $5\kappa(=1)$.}
\label{fig1}
\end{figure}

Figure~\ref{fig1} illustrates variation of the steady-state degree of coherence~$|\gamma_{(R,L)}|$ with $\beta$ for several different values of~$\alpha_{k}$. It is seen that after the interaction with the finite size atomic ensemble, there is a non-zero mutual coherence between the counter-propagating cavity modes. The coherence increases with $\alpha_{k}$ and the modes become perfectly correlated, $|\gamma_{(R,L)}|\rightarrow 1$ as $\alpha_{k}$ tends towards unity. Moreover, the coherence becomes less sensitive to $\beta$ as $\alpha_{k}$ increases. In addition, for $\alpha_{k}\approx 1$ the coherence attains its maximal value of $|\gamma_{(R,L)}|=1$ independent of $\beta$. Notice, that the threshold value of the coupling  strength at which $|\gamma_{(R,L)}|$ approaches unity, shifts towards smaller $\beta$ as $\alpha_{k}$ increases. The threshold value of $\beta$ corresponds to a critical value of $\beta$.

We close this section by evaluating the degree of coherence in the case of $\alpha_{k} =1$, that is, when the size of the atomic ensemble is much smaller than the resonant wavelength of the cavity modes, i.e., when $2\vec{k}_{c}\cdot \vec{r}_{j}\ll 1$. To consider that limit, we can return to the effective Hamiltonian and note that with $\alpha_{k}$ increasing to the value unity, the collective mode $d_{2}$ becomes decoupled from the cavity field mode $a_{2}$. In this case, the effective Hamiltonian reduces to that of a single-mode Dicke system involving only the field mode $a_{1}$ and the collective mode $d_{1}$. The modes $a_{1}$ and $d_{1}$ undergo the time evolution, whereas the modes $a_{2}$ and $d_{2}$ remain constant in time that they retain their initial values for all times. Although the behaviours of the system  effectively as a single-mode Dicke system, it, in fact, involves two modes since the mode $a_{1}$ is a superposition of the cavity counter-propagating modes.

The steady-state solutions (\ref{B1}) are not valid for $\alpha_{k}=1$. However, almost all mode correlation functions can be obtained from~(\ref{B1}) by putting $\lambda_{2}=0$ except of $\langle\hat{a}^{\dagger}_{2}\hat{a}_{2}\rangle$ and $\langle\hat{d}^{\dagger}_{2}\hat{d}_{2}\rangle$. These two correlation functions are constants of motion when $\alpha_{k}=1$. As a result, the modes do not evolve in time, they retain their initial values for all times $t$. Thus, if initially the modes were unpopulated, they will stay unpopulated for all times. This somewhat unusual result is a consequence of the fact that the modes are totally decoupled from the applied field. The immediate consequence of the decoupling of the modes $\hat{a}_{2}$ and $\hat{d}_{2}$ from the field is the appearance of the perfect correlation between the cavity counter-propagating modes.
It is easy seen from (\ref{e32a}) and~(\ref{e33a}) that in this case, the degree of the first-order coherence $|\gamma_{(R,L)}|=1$ irrespective of the parameters involved. This result has a simple interpretation, the mode $\hat{a}_{1}$ that can be prepared in an arbitrary state, is a linear superposition of the cavity counter-propagating modes that enter with equal weights. Therefore, both modes are always equally prepared, so that cannot be resolved, which is reflected in the coherence equal to unity.

\subsection{Second-order coherence of the cavity modes}\label{sec3b}

We now consider the second-order correlation functions of the fields of individual modes and of two different modes. We are particularly interested in the correlations in the cavity counter-propagating modes, represented by the operators $\hat{a}_{R}$ and $\hat{a}_{L}$, and between these modes. The correlations are determined by $\langle \hat{a}_{R}^{\dag}\hat{a}_{R}^{\dag}\hat{a}_{R}\hat{a}_{R}\rangle$, $\langle \hat{a}_{L}^{\dag}\hat{a}_{L}^{\dag}\hat{a}_{L}\hat{a}_{L}\rangle$, and $\langle \hat{a}_{R}^{\dag}\hat{a}_{L}^{\dag}\hat{a}_{R}\hat{a}_{L}\rangle$, respectively. More specifically, correlation functions describe the photon statistics of the field of the individual modes, and the cross correlations between photons from two different modes. As we shall see below, the second-order correlation functions, especially the cross correlation function $\langle \hat{a}_{R}^{\dag}\hat{a}_{L}^{\dag}\hat{a}_{R}\hat{a}_{L}\rangle$ can be used to determine nonclassical effects and entanglement between the modes~\cite{sv05,hz07}.

We shall consider normalized correlation functions and assume that the cavity modes are Gaussian-state modes that they obey the moment-factorization rules of a Gaussian random variable~\cite{gz00}. This is to be expected, since the collective bosonic modes are the sum of a large but finite number of atoms $N$, whose fluctuations have been supposed to be statistically independent. Therefore, the Gaussian form of the collective bosonic modes is expected to be reflected in a Gaussian form of the other modes. With this condition, the normalized second-order correlation functions can be written as
\begin{eqnarray}
g^{(2)}_{RR} &= \frac{\langle \hat{a}_{R}^{\dag}\hat{a}_{R}^{\dag}\hat{a}_{R}\hat{a}_{R}\rangle}{\langle \hat{a}_{R}^{\dag}\hat{a}_{R}\rangle^{2}} = 2 + \left|\eta_{(R,R)}\right|^{2}  ,\nonumber\\
g^{(2)}_{LL} &= \frac{\langle \hat{a}_{L}^{\dag}\hat{a}_{L}^{\dag}\hat{a}_{L}\hat{a}_{L}\rangle}{\langle \hat{a}_{L}^{\dag}\hat{a}_{L}\rangle^{2}} = 2 + \left|\eta_{(L,L)}\right|^{2}  ,\nonumber\\
g^{(2)}_{RL} &= \frac{\langle \hat{a}_{R}^{\dag}\hat{a}_{L}^{\dag}\hat{a}_{R}\hat{a}_{L}\rangle}{\langle \hat{a}_{R}^{\dag}\hat{a}_{R}\rangle\langle \hat{a}_{L}^{\dag}\hat{a}_{L}\rangle} = 1 + \left|\gamma_{(R,L)}\right|^{2} +\left|\eta_{(R,L)}\right|^{2} ,\label{eq38}
\end{eqnarray}
where $\left|\gamma_{(R,L)}\right|$ is the degree of the first-order coherence, defined in equation~(\ref{eq32}), and
\begin{eqnarray}
\left|\eta_{(L,L)}\right| &= \left|\eta_{(R,R)}\right|=\frac{\left|\langle \hat{a}_{R}\hat{a}_{R}\rangle\right|}{\langle \hat{a}_{R}^{\dag}\hat{a}_{R}\rangle}= \frac{\left|\langle \hat{a}^2_{1}\rangle + \langle \hat{a}^2_{2}\rangle\right|}{\langle \hat{a}_{1}^{\dag}\hat{a}_{1}\rangle + \langle \hat{a}_{2}^{\dag}\hat{a}_{2}\rangle} ,\nonumber\\
\left|\eta_{(R,L)}\right| &= \frac{\left|\langle \hat{a}_{R}\hat{a}_{L}\rangle\right|}{\sqrt{\langle \hat{a}_{R}^{\dag}\hat{a}_{R}\rangle\langle \hat{a}_{L}^{\dag}\hat{a}_{L}\rangle}}=\frac{\left|\langle \hat{a}^2_{1}\rangle - \langle \hat{a}^2_{2}\rangle\right|}{\langle \hat{a}_{1}^{\dag}\hat{a}_{1}\rangle + \langle \hat{a}_{2}^{\dag}\hat{a}_{2}\rangle} ,\label{eq39}
\end{eqnarray}
are degrees of the so-called "anomalous" coherence~\cite{ks80,lo84,ag86,hr87,ft88}. We see that the most important contribution to the second-order correlation functions comes from the anomalous coherence functions. These relations also show that $g^{(2)}_{RR}\geq 2$ and $g^{(2)}_{LL}\geq 2$, which means that photons emitted in the same direction, $R$ or $L$, are always strongly correlated. This is a reflection of the Gaussian statistics of the cavity modes that the four-order moments $\langle \hat{a}_{R}^{\dag}\hat{a}_{R}^{\dag}\hat{a}_{R}\hat{a}_{R}\rangle$ and $\langle \hat{a}_{L}^{\dag}\hat{a}_{L}^{\dag}\hat{a}_{L}\hat{a}_{L}\rangle$ factorize into $\left|\langle \hat{a}_{R}\hat{a}_{R}\rangle\right|^{2}+2\langle \hat{a}_{R}^{\dag}\hat{a}_{R}\rangle^{2}$ and $\left|\langle \hat{a}_{L}\hat{a}_{L}\rangle\right|^{2}+2\langle \hat{a}_{L}^{\dag}\hat{a}_{L}\rangle^{2}$, respectively.
No such factorisation is possible for a non-Gaussian statistics of the modes that may result in photon antibunching effect~\cite{kd77,dm78,ch82,dw87,fost00} of $g^{(2)}_{RR}<1$ or $g^{(2)}_{LL}<1$.

The inter-mode second-order correlation function is the sum of contributions $|\gamma_{(R,L)}|^{2}$ and $|\eta_{(R,L)}|^{2}$, which indicates that the correlation of photons emitted in opposite directions depends on two kinds of coherence, the mutual first-order coherence and mutual anomalous coherence.

Let us examine the dependence of the correlation functions on the ensemble size parameter $\alpha_{k}$. For purposes of an explicit analytical analysis, it is somewhat more convenient to rewrite~(\ref{eq32}) in terms of the superposition cavity modes~$1$ and $2$. With the help of~(\ref{a12}), we arrive at the expressions
\begin{eqnarray}
\left|\gamma_{(R,L)}\right| &= \frac{\left|\langle \hat{a}_{1}^{\dag}\hat{a}_{1}\rangle - \langle \hat{a}_{2}^{\dag}\hat{a}_{2}\rangle\right|}{\langle \hat{a}_{1}^{\dag}\hat{a}_{1}\rangle + \langle \hat{a}_{2}^{\dag}\hat{a}_{2}\rangle} .\label{eq40}
\end{eqnarray}
We observe that the inter-mode coherence function $|\gamma_{(R,L)}|$ depends on the difference between the number of photons in the superposition modes, whereas the inter-mode coherence function $|\eta_{(R,L)}|$ depends on the difference between the anomalous coherence functions of the modes. Thus, some kind of asymmetry between the superposition modes is needed to create the coherence between the cavity counter-propagating modes. The coherence~(\ref{eq39}) and (\ref{eq40}) can be readily evaluated using the steady-state solutions~(\ref{B1}).

Consider first the correlation functions in the thermodynamic limit, in which case $\alpha_{k}=0$. From the steady-state solutions, (\ref{B1}), it follows that in the limit of $\alpha_{k}=0$, $\langle \hat{a}_{1}^{\dag}\hat{a}_{1}\rangle = \langle \hat{a}_{2}^{\dag}\hat{a}_{2}\rangle$, $\langle \hat{a}^2_{1}\rangle = -\langle\hat{a}^2_{2}\rangle$, and $|\langle \hat{a}^2_{1}\rangle| = \langle \hat{a}_{1}^{\dag}\hat{a}_{1}\rangle$, so that
\begin{eqnarray}
\left|\eta_{(R,R)}\right| = \left|\eta_{(L,L)}\right| = \left|\gamma_{(R,L)}\right| = 0 ,\qquad \left|\eta_{(R,L)}\right| = 1 ,\label{eq41}
\end{eqnarray}
and from (\ref{eq38}), we immediately obtain that
\begin{eqnarray}
g^{(2)}_{RR} = g^{(2)}_{LL} = 2 ,\qquad {\rm and} \qquad g^{(2)}_{RL} = 2 .
\end{eqnarray}
These results show that in the thermodynamic limit the cavity modes and the correlation between the modes exhibit correlations characteristic of a thermal field. It is, of course, a reflection of the fact that the system operates below the threshold where the modes are in thermal states. This is the kind of behavior that is expected for the cavity modes. One could argue that the same circumstances apply for the presence of the correlations between the modes. However, the circumstances for the second-order correlations $g^{(2)}_{RL} = 2$ are different. The source of the correlation between the modes is not in the thermal fluctuations, as it takes place in the well-known Hanbury-Brown-Twiss effect, but is in the impossibility to distinguish from which mode each of the two photons came. This is represented by the anomalous coherence $|\eta_{(R,L)}|$, and for this reason, we could call this effect as an {\it anomalous} Hanbury-Brown-Twiss effect.

\begin{figure}[hbt]
\begin{center}
\begin{tabular}{c}
\includegraphics[height=6cm,width=0.47\columnwidth]{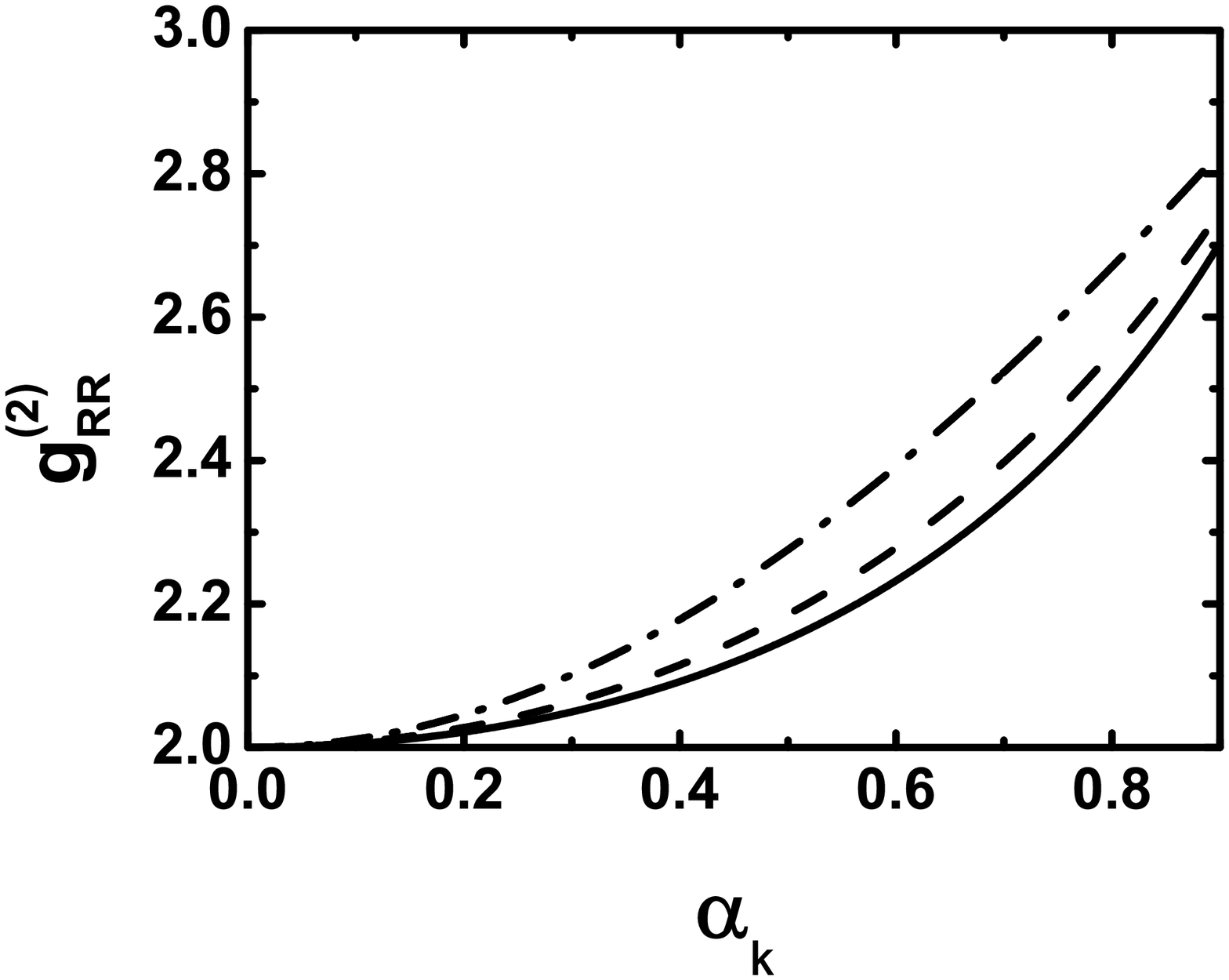}
\includegraphics[height=6cm,width=0.47\columnwidth]{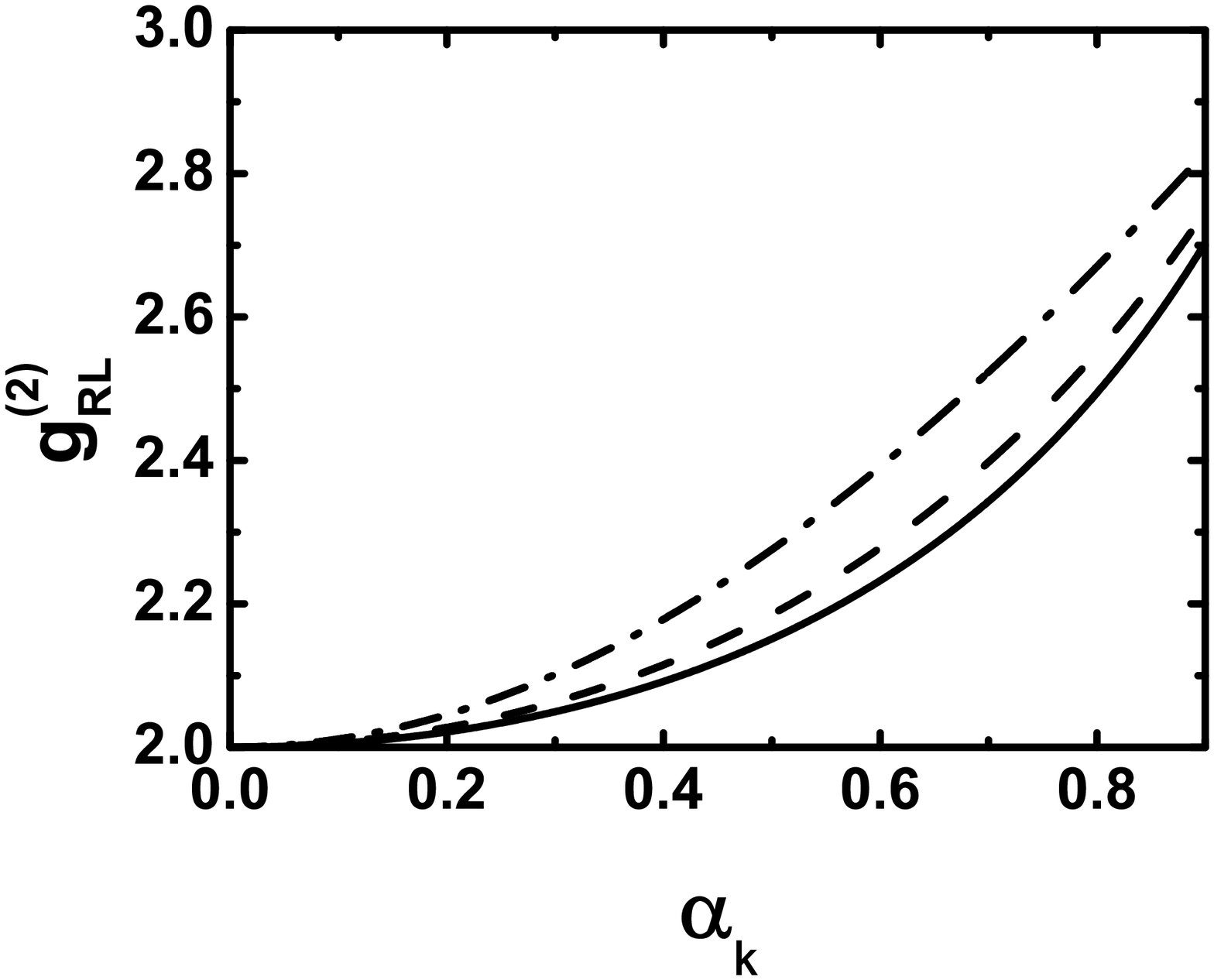}
\end{tabular}
\end{center}
\caption[nsa]{The stationary second-order correlation functions $g^{(2)}_{RR}$ (left frame) and $g^{(2)}_{RL}$ (right frame) plotted as a function of the finite-size parameter $\alpha_{k}$ for $\omega_0=\omega=1,\delta=0.1 \pi,\kappa=0.2$, and different values of the coupling strength $\beta$: $\beta=0.1$ (solid line), $\beta =0.2$ (dashed line), and $\beta =0.3$ (dashed-dotted line).}
\label{fig2}
\end{figure}

We have already seen that in the thermodynamic limit only the mutual anomalous coherence $\left|\eta_{(R,L)}\right|$ is different from zero and, in fact, attains its maximal value of unity. It is clear by inspection of (\ref{eq39}) that an asymmetry between the superposition modes is required to get all of the coherence different from zero. It is easily verified from (\ref{B1}) that the required asymmetry is provided by the ensemble size parameter $\alpha_{k}$.
In this case, the second-order correlations can be larger than that for the thermal field. This is known in the literature as a super-bunching effect~\cite{pl86,jy86,me05,ak07}. Hence, $\alpha_{k}\neq 0$ is the general condition for the super-bunching effect. The variation of the correlation functions with $\alpha_{k}$ for several different values of $\beta$ is illustrated in figure~\ref{fig2}. It is evident that the finite-size effects enhance the correlations between photons emitted in the same as well as in the opposite directions.

The largest value of the correlations is achieved when $\alpha_{k}=1$, in this case $g^{(2)}_{RR}=g^{(2)}_{LL}=g^{(2)}_{RL}=3$. This value is the border value between classical and nonclassical Gaussian states~\cite{jy86,sw07}.
We may conclude that the output cavity modes behaviour as an unusual classically correlated reservoir which exhibits strong classical correlations simultaneously inside the individual modes and also between the modes. Typical sources of correlated beams, such as optical parametric oscillators exhibit correlations stronger than that of a thermal field only between the modes.

One can also notice from the figure~\ref{fig2} that the correlations functions $g^{(2)}_{RR}$ and $g^{(2)}_{LL}$ behaviour similarly to the mutual correlation function $g^{(2)}_{RL}$. However, there is a relation between the correlation functions, given by the Cauchy-Schwartz inequality~\cite{mw95}
\begin{eqnarray}
\chi_{RL} = \frac{g^{(2)}_{RR}g^{(2)}_{LL}}{\left[g^{(2)}_{RL}\right]^{2}} \geq 1 ,\label{eq43}
\end{eqnarray}
which says that the cross correlations between photons from the two different cavity modes are smaller than the correlation between photons of the individual modes. An interesting question arises whether the Cauchy-Schwartz inequality can be violated in the system.
\begin{figure}[hbt]
\begin{center}
\begin{tabular}{c}
\includegraphics[height=8cm,width=0.75\columnwidth]{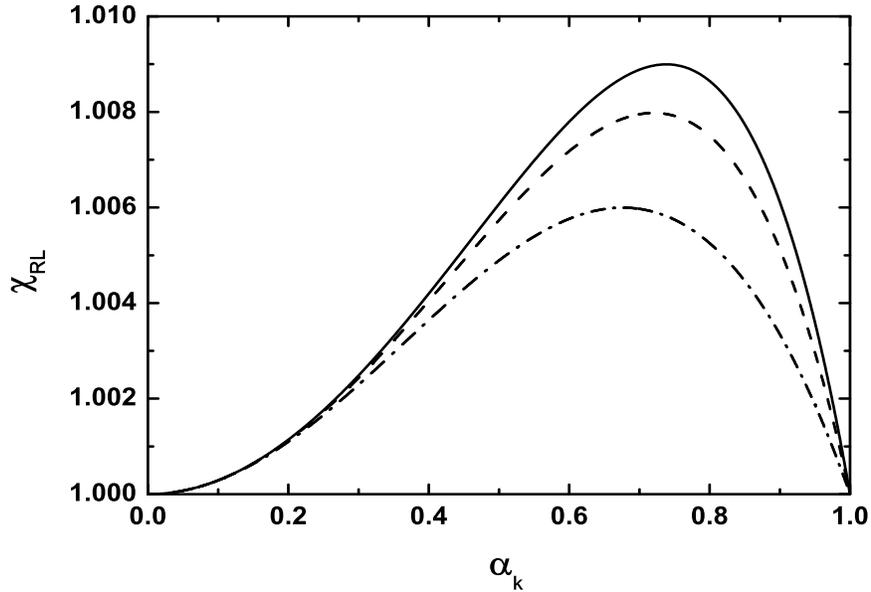}
\end{tabular}
\end{center}
\caption[nsa]{Variation of the Cauchy-Schwartz parameter $\chi_{RL}$ with the finite-size parameter $\alpha_{k}$ for $\omega_0=\omega=1,\delta=0.1 \pi,\kappa=0.2$, and different values of the coupling strength $\beta$: $\beta=0.1$ (solid line), $\beta =0.2$ (dashed line), and $\beta =0.3$ (dashed-dotted line).}
\label{fig3}
\end{figure}

Figure~\ref{fig3}  illustrates the variation of the Cauchy-Schwartz parameter $\chi_{RL}$ with $\alpha_{k}$. We see that even when the correlations functions $g^{(2)}_{RR}$ and $g^{(2)}_{LL}$ behaviour similar to the mutual correlation function $g^{(2)}_{RL}$, the Cauchy-Schwartz parameter varies with $\alpha_{k}$. It is apparent that $\chi_{RL}$ is always more than or equal to unity for all $\alpha_{k}$, with equality at $\alpha_{k}=0$ and $\alpha_{k}=1$, indicating that the Cauchy-Schwartz inequality is not violated. Thus, the finite-size effects create strong correlations between the cavity modes but do not allow the relation~(\ref{eq43}) to be violated.

It is not difficult to show from (\ref{eq38}) and (\ref{eq39}) that the inequality~(\ref{eq43}) is equivalent to the inequality $\left|\eta_{(R,L)}\right|\leq 1$, that for the Cauchy-Schwartz inequality to be satisfied, the mutual anomalous coherence must be smaller than unity. Thus, for a violation of the Cauchy-Schwartz inequality it is necessary that the degree of the mutual anomalous coherence to be larger than unity. It is worth noting that such values can be achieved only by a quantum field~\cite{mw95,zw91,hz07}.
\begin{figure}[hbt]
\begin{tabular}{c}
\includegraphics[height=6cm,width=0.48\columnwidth]{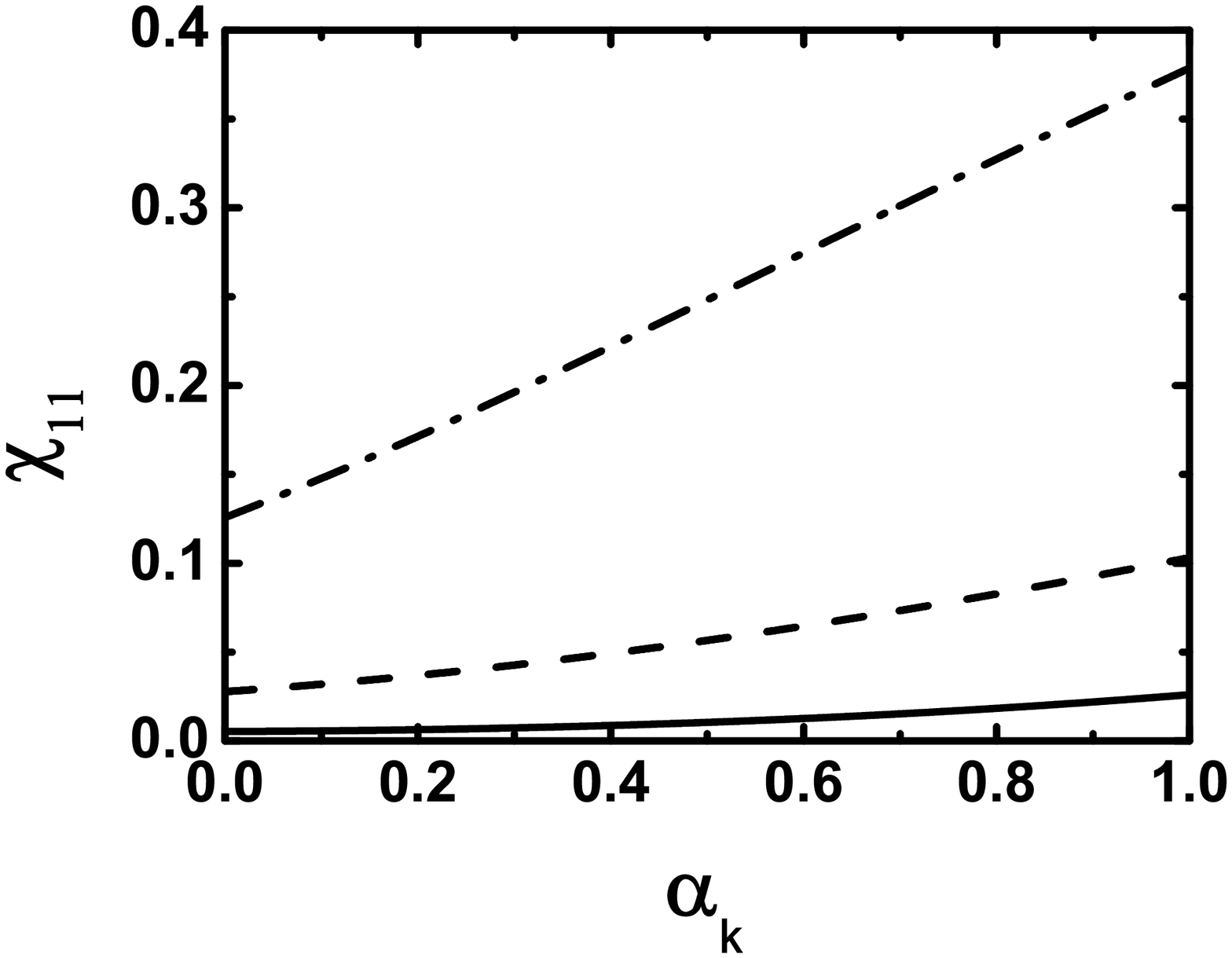}
\includegraphics[height=6cm,width=0.50\columnwidth]{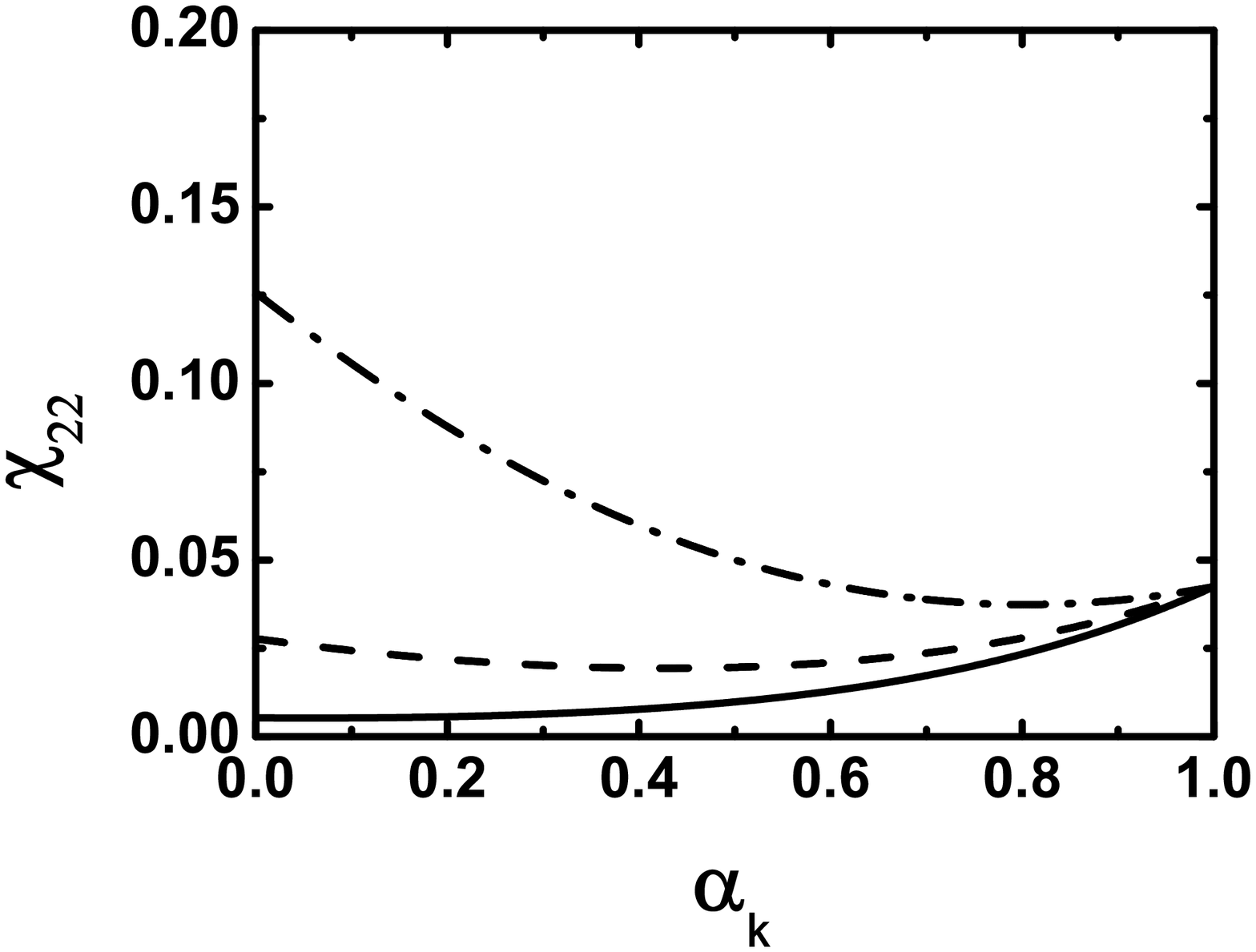}
\end{tabular}
\caption[nsa]{Variation of the Cauchy-Schwartz parameters $\chi_{11}$ (left frame) and $\chi_{22}$ (right frame) with the finite-size parameter $\alpha_{k}$ for $\omega_0=\omega=1,\delta=0.1 \pi,\kappa=0.2$, and different values of the coupling strength $\beta$: $\beta=0.1$ (solid line), $\beta =0.2$ (dashed line), and $\beta =0.3$ (dashed-dotted line).}
\label{fig4}
\end{figure}

Although the Cauchy-Schwartz inequality is not violated between the cavity modes, it may be violated between other modes. Figure~\ref{fig4} shows Cauchy-Schwarz parameters $\chi_{11}$ and~$\chi_{22}$ defined as
\begin{eqnarray}
\chi_{11} = \frac{\langle \hat{a}_{1}^{\dag 2}\hat{a}_{1}^{2}\rangle\langle \hat{d}_{1}^{\dag 2}\hat{d}_{1}^{2}\rangle}{\langle \hat{a}_{1}^{\dag}\hat{d}_{1}^{\dag}\hat{a}_{1}\hat{d}_{1}\rangle^{2}} ,\qquad \chi_{22} = \frac{\langle \hat{a}_{2}^{\dag 2}\hat{a}_{2}^{2}\rangle\langle \hat{d}_{2}^{\dag 2}\hat{d}_{2}^{2}\rangle}{\langle \hat{a}_{2}^{\dag}\hat{d}_{2}^{\dag}\hat{a}_{2}\hat{d}_{2}\rangle^{2}} ,\label{eq44}
\end{eqnarray}
which provide measures of the second-order correlations between photons from two superposition modes $(\hat{a}_{1}$ and $\hat{d}_{1})$, and from other two superposition modes $(\hat{a}_{2}$ and $\hat{d}_{2})$, respectively. The correlations are said to violate the Cauchy-Schwartz inequality if $\chi_{ii}\, (i=1,2)$ is smaller than unity. It is seen  that the Cauchy-Schwartz inequality is violated for both pairs of the modes indicating a strong nonclassical correlation between the superposition modes. These violations exist even for $\alpha_{k}=0$ and decrease with increasing pumping strength $\beta$. The physical reason for the violation of the Cauchy-Schwartz inequality can be traced to nonlinear processes that are known to produce quantum effects in the interaction between bosonic modes~\cite{df04}. It is evident from the effective Hamiltonian~(\ref{he3}) that such processes exist in the system. There is a nonlinear coupling between modes~$\hat{a}_{1}$ and~$\hat{d}_{1}$, and between modes $\hat{a}_{2}$ and $\hat{d}_{2}$. The effect of these nonlinear couplings is to produce nonzero anomalous correlation functions that are responsible for enhanced inter-mode correlations.

In summary of this section, we have found that the role played by the finite-size effects in the second-order correlations is principally to create correlations which are larger than that achievable with thermal fields. However, there is a limitation on the values of the second-order correlations that could be created by the finite-size effects. The second-order correlation functions vary with the finite-size parameter $\alpha_{k}$ from $g^{(2)}_{ij} =2$ for $\alpha_{k}=0$ to the maximum of $g^{(2)}_{ij} =3$ for $\alpha_{k}=1$, which is achieved when the dimensions of the atomic ensemble are much smaller that the resonant wavelength. These results show that the total field emerging from the cavity is a classical but strongly correlated thermal field. It was found that quantum effects such as the violation of the Cauchy-Schwartz inequality can be created between the superposition modes. Unfortunately, the quite large violations of the Cauchy-Schwartz inequality for the superposition modes do not lead to violation of the Cauchy-Schwartz inequality for the correlations between the cavity counter-propagating modes.

\subsection{Entanglement and two-mode squeezing spectra of the output cavity modes}\label{sec3d}

Since the finite-size effects create first-order coherence and the second-order correlations between the modes, there actually could be two-mode squeezing and entanglement between the modes associated with a nonlinear coupling between the modes  as well. An inspection of the effective Hamiltonian~(\ref{he3}) reveals that a nonlinear coupling actually exists only between modes $\hat{a}_{1}$ and $\hat{d}_{1}$, and between modes $\hat{a}_{2}$ and $\hat{d}_{2}$. Thus, the modes $(\hat{a}_{1},\hat{d}_{1})$ and $(\hat{a}_{2},\hat{d}_{2})$ could be entangled between themselves. This suggests that the other pairs of the modes cannot be entangled. We now examine the possibility to create two-mode squeezing and entanglement between the cavity counter-propagating modes $a_L$ and $a_R$ and how these effects could depend on the finite-size parameter $\alpha_{k}$.

In order to find if entanglement and two-mode squeezing can be created between the cavity modes $a_{R}$ and $a_{L}$ modes, we introduce the position and momentum operators for the annihilation operators of the superposition modes, $\hat{a}_{1}$ and $\hat{a}_{2}$, which can be defined as
\begin{eqnarray}
X_{j}^{\theta} &= \frac{1}{\sqrt{2}}\left(a_{j}{\rm e}^{i\theta} +a^{\dagger}_{j}{\rm e}^{-i\theta}\right),\nonumber\\
P_{j}^{\theta} &= \frac{i}{\sqrt{2}}\left(a_{j}^{\dag}{\rm e}^{-i\theta} -a_{j}{\rm e}^{i\theta}\right) ,\, j=1,2 ,
\end{eqnarray}
where $\theta$ is the quadrature phase.

To see if an entanglement exists between the cavity counter-propagating modes, $a_R$ and $a_L$, we use a condition based on the two-mode squeezing, proposed by van Loock and Furusawa~\cite{lf03}. By use of the mode transformations for the modes $a_L$, $a_R$ $a_1$, and $a_2$, (\ref{a12}), the sufficient condition for the entanglement between the two cavity modes $a_R$ and $a_L$ is of the form~\cite{lf03}
\begin{eqnarray}
\left\langle:\left(\Delta X_{1}^{\theta}\right)^{2}:\right\rangle + \left\langle :\left(\Delta P_{2}^{\theta}\right)^{2}:\right\rangle < 0 ,\label{lf}
\end{eqnarray}
where the normally ordered variances are given by
\begin{eqnarray}
\left\langle:\Delta \left(\hat{X}^\theta_{1}\right)^{2}:\right\rangle = \langle a_1^\dag a_1\rangle \cos^2(\theta+\varphi_1)  ,\nonumber\\
\left\langle:\Delta \left(\hat{P}^\theta_{2}\right)^{2}:\right\rangle = \langle a_2^\dag a_2\rangle \cos^2(\theta+\varphi_2) ,\label{e58}
\end{eqnarray}
with $\varphi_j=\arctan(\kappa/\omega_{j})$. Here, the double colon $::$ stands for the normal ordering of the operators.

When we evaluate the normally ordered variances~(\ref{e58}) using the steady state solutions~(\ref{B1}), we then easily find that $\langle:  \left( \Delta X_{1}^{\theta}\right)^{2}  :\rangle + \langle : \left( \Delta P_{2}^{\theta}\right)^{2} :\rangle > 0$ for any $\theta$. Thus, there is no squeezing between the cavity modes. Equivalently, the cavity modes are separable. We may conclude that any measurable criterion predicts no squeezing and thus no entanglement of the total field of the counter-propagating cavity modes $a_L$ and $a_R$.
On the other hand, from (\ref{arl}) and~(\ref{eq39}), we see that $\langle \hat{a}_{R}^{\dag}\hat{a}_{L}\rangle\neq0$, and $\langle \hat{a}_{R}\hat{a}_{L}\rangle\neq 0$ when $\alpha_k\neq0$. This means that the modes are correlated and the strength of correlation depends on the values of $\alpha_k$. Hence, the modes are correlated but not strong enough to be squeezed (entangled). This conclusion agrees with our previous findings that the modes are correlated to the degree of $g^{(2)}_{RL}=3$, which is the border value between classical and nonclassical Gaussian states.  This means that the correlations created in the pairs $(\hat{a}_{1},\hat{d}_{1})$ and $(\hat{a}_{2},\hat{d}_{2})$ can be transferred to the cavity modes, but are not strong enough to entangle the modes.

We stress that the calculated correlations corresponded to that of the total cavity field. It is well known that in some situations there is no squeezing in the total field, but there could be squeezing between spectral components of the field~\cite{ls82,oh87}. In other words, the question of whether the total output field is squeezed (entangled) may be irrelevant to the problem of obtaining large amount of squeezing (entanglement) at some particular spectral frequency. Nevertheless, we shall show that a strong squeezing (entanglement) exists between spectral components of the output cavity fields.

We now use the frequency dependent solutions for the cavity modes and the relations between the input and output fields~\cite{cg84}
\begin{eqnarray}
\tilde{a}_{j}^{out}(\nu) = \sqrt{2\kappa}\,\tilde{a}_{j}(\nu)-\tilde{a}_{j}^{in}(\nu) ,\label{lambqv}
\end{eqnarray}
where $\tilde{a}_{j}(\nu)$ are the intracavity field operators and $\tilde{a}_{j}^{in}(\nu)$ are the input noise operators. 
We shall use the frequency dependent operators to calculate the measurable spectra of the output fields transmitted by one of the cavity mirror with decay constant $\kappa$. Let~$\nu$ is the frequency of the output clockwise $(R)$ mode and $\nu^{\prime}$ is the frequency of the output anti-clockwise $(L)$ mode. We may introduce finite frequency intervals $\delta\nu$ and $\delta\nu^{\prime}$ defined~as
\begin{eqnarray}
\delta\nu = \bar{\omega}_{l} -\nu , \quad \delta\nu^{\prime} =-\bar{\omega}_{l} +\nu^{\prime} ,
\end{eqnarray}
such that at $\delta\nu =\delta\nu^{\prime}$ the modes are symmetrically located about $2\bar{\omega}_{l}$, i.e. $\nu +\nu^{\prime}=2\bar{\omega}_{l}$.

We consider spectral distributions of the variances
\begin{eqnarray}
\left\langle :\left(\Delta \tilde{X}_{1}^{\theta}(\nu)\right)^2 :\right\rangle + \left\langle :\left(\Delta \tilde{P}_{2}^{\theta}(\nu^{\prime})\right)^{2}:\right\rangle = S(\nu,\theta)\delta(2\bar{\omega}_{l}-\nu -\nu^{\prime}) ,\label{Delta}
\end{eqnarray}
where $\tilde{X}_{1}^{\theta}(\nu)$ and $\tilde{P}_{2}^{\theta}(\nu^{\prime})$ are Fourier transforms of the quadrature phase operators of the superpositions $\hat{a}_{1}$ and $\hat{a}_{2}$ of the output modes, defined~as
\begin{eqnarray}
\tilde{X}_{1}^{\theta}(\nu) &=& \frac{1}{\sqrt{2}}\left[\tilde{a}_{1}^{out}(\nu){\rm e}^{i\theta}
+\tilde{a}^{out \dagger}_{1}(-\nu){\rm e}^{-i\theta}\right] ,\nonumber\\
\tilde{P}_{2}^{\theta}(\nu^{\prime}) &=& \frac{i}{\sqrt{2}}\left[\tilde{a}_{2}^{out \dagger}(-\nu^{\prime}){\rm e}^{-i\theta}
-\tilde{a}^{out}_{2}(\nu^{\prime}){\rm e}^{i\theta}\right] .
\end{eqnarray}

We also consider the spectral distribution of the logarithmic negativity criterion that is known as the necessary and sufficient condition for entanglement of two-mode Gaussian states~\cite{si00,as04}
\begin{eqnarray}
E_{n}(\nu)  = {\rm max}\left\{0,-\log_{2}\left[2V_{s}(\nu)\right]\right\} ,
\end{eqnarray}
where $V_{s}(\nu)$ is the smallest sympletic eigenvalue of the partially transposed covariance matrix of the output field. We shall compare the criterion with the two-mode squeezing criterion to quantify squeezing as an alternative necessary and sufficient condition for entanglement~\cite{gv10}. The advantage of the two-mode squeezing criterion over the negativity is that the former can be directly measured in experiments whereas the later can be inferred from the reconstruction of the density matrix of the system.

To evaluate $E_{n}(\nu)$, that describe entanglement of a two-mode output Gaussian state, we use Wigner characteristic function
\begin{eqnarray}
\chi(\xi_{\hat{a}_{R}},\xi_{\hat{a}_{L}})
= \exp\left(-\frac{1}{2} \xi V(\nu) \xi^T \right) ,\label{xzeta}
\end{eqnarray}
where $\xi =(\xi^\ast_{\hat{a}_{R}},\xi_{\hat{a}_{R}},\xi^\ast_{\hat{a}_{L}},\xi_{\hat{a}_{L}})$ is a vector of complex variables, $\xi^T$ is the transposed for of $ \xi$, and $V(\nu)$ is the covariance matrix of the form
\begin{eqnarray}
V(\nu) = \left(\begin{array}{cccc}
f_{1}(\nu)&f_{2}(\nu) &f_{3}(\nu)&f_{4}(\nu)\\
f^\ast_{2}(\nu)&f_{1}(\nu)&f^\ast_{4}(\nu)&f_{3}(\nu)\\
f_{3}(\nu)&f_{4}(\nu)&f_{1}(\nu)&f_{2}(\nu)\\
f^\ast_{4}(\nu)&f_{3}(\nu)&f^\ast_{2}(\nu)&f_{1}(\nu)
\end{array}\right) .
\label{vef}
\end{eqnarray}
with
\begin{eqnarray}
&&f_{1}(\nu) - \frac{1}{2} = \langle \tilde{a}^{out \dag}_{R}(\nu),\tilde{a}^{out}_{R}(\nu^\prime)\rangle =\langle \tilde{a}^{out \dag}_{L}(\nu),\tilde{a}^{out}_{L}(\nu^\prime)\rangle \nonumber\\
&&= \kappa \left[\frac{M^\ast_{12}(\nu)M_{12}(\nu^\prime)}{D^\ast_1(\nu)D_1(\nu^\prime)} + \frac{M^\ast_{22}(\nu)M_{22}(\nu^\prime)}{D^\ast_2(\nu)D_2(\nu^\prime)}\right]\delta(\nu-\nu^\prime) ,\nonumber\\
&&f_{2}(\nu) = \langle \tilde{a}^{out}_{R}(\nu),\tilde{a}^{out}_{R}(\nu^\prime)\rangle\nonumber\\
&& =\kappa\!\left\{\frac{\left[M_{11}(\nu)-\frac{D_1(\nu)}{\sqrt{2 \kappa}}\right]\!M_{12}(\nu^\prime)}{D_1(\nu)D_1(\nu^\prime)}+\frac{\left[M_{21}(\nu)-\frac{D_2(\nu)}{\sqrt{2 \kappa}}\!\right]\!M_{22}(\nu^\prime)}{D_2(\nu)D_2(\nu^\prime)}\right\}\!\delta(2\bar{\omega}_{l}-\nu -\nu^{\prime}) ,\nonumber\\
&&f_{3}(\nu) = \langle \tilde{a}^{out \dag}_{L}(\nu),\tilde{a}^{out}_{R}(\nu^\prime)\rangle \nonumber\\
&&= \kappa \left[\frac{M^\ast_{12}(\nu)M_{12}(\nu^\prime)}{D^\ast_1(\nu)D_1(\nu^\prime)} -\frac{M^\ast_{22}(\nu)M_{22}(\nu^\prime)}{D^\ast_2(\nu)D_2(\nu^\prime)}\right]\delta(\nu-\nu^\prime) ,\\
&&f_{4}(\nu) = \langle \tilde{a}^{out}_{L}(\nu),\tilde{a}^{out}_{R}(\nu^\prime)\rangle \nonumber\\
&&= \kappa\!\left\{\!\frac{\left[\!M_{11}(\nu)-\frac{D_1(\nu)}{\sqrt{2 \kappa}}\!\right]\!M_{12}(\nu^\prime)}{D_1(\nu)D_1(\nu^\prime)} - \frac{\left[\!M_{21}(\nu)-\frac{D_2(\nu)}{\sqrt{2 \kappa}}\!\right]\!M_{22}(\nu^\prime)}{D_2(\nu)D_2(\nu^\prime)}\!\right\}\!\delta(2\bar{\omega}_{l}-\nu -\nu^{\prime}) .\nonumber
\end{eqnarray}
This shows that $f_{2}(\nu), f_{3}(\nu)$ and $f_{4}(\nu)$ determine the correlation between the two output cavity modes.
\begin{figure}[hbt]
\begin{center}
\begin{tabular}{c}
\includegraphics[height=12cm,width=0.9\columnwidth]{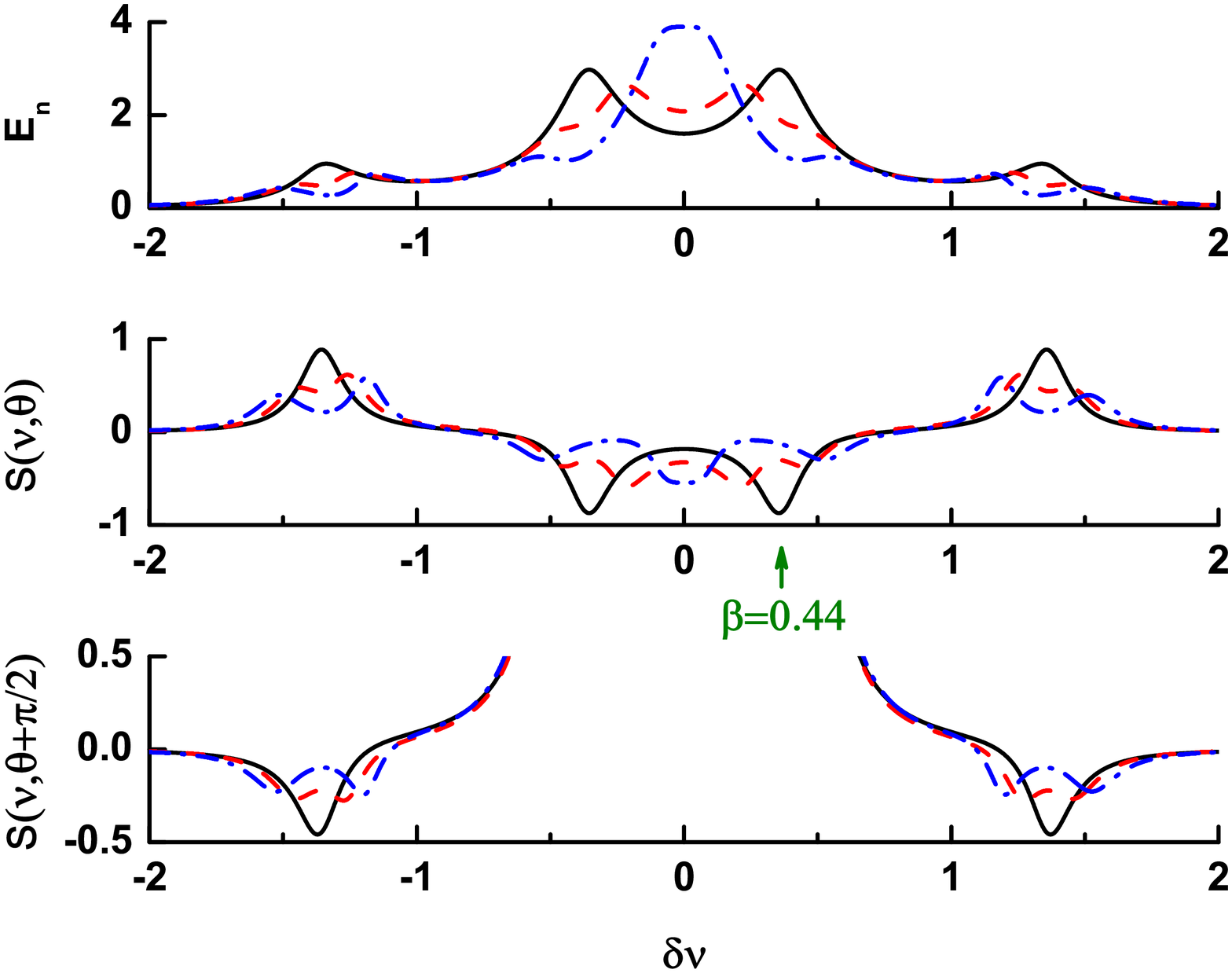}
\end{tabular}
\end{center}
\caption[nsa]{\label{fig6} The variation of the logarithmic negativity $E_{n}(\nu)$ and the variances $S(\nu,\theta)$ with the frequency interval $\delta\nu =\bar{\omega}_{l}-\nu$ for $\omega_0=\omega=1,\delta=0.1\pi,\beta =0.44$, $\kappa=0.2$ and several different values of $\alpha_{k}$: $\alpha_{k}=0,\theta=1.6856$ (solid line), $\alpha_{k}=0.3,\theta=1.6518$ (dashed line), $\alpha_{k}=0.5,\theta=1.6676$ (dotted line). The arrow indicates frequency of the generalized Rabi frequency $\beta$.}
\end{figure}

Having the covariance matrix, we may discuss in detail the establishment of entanglement between two output cavity modes. We shall be particularly interested in the role of the finite-size effects on the output entanglement of the two counter-propagating cavity modes.

Figure~\ref{fig6} shows the spectral distribution of the logarithmic negativity and the variances of the output fields for different~$\alpha_{k}$. We also vary the phase $\theta$ due to varying of the optimal squeezing with an increasing $\alpha_{k}$. First, we note that independent of $\alpha_{k}$, it is possible for $S(\nu,\theta)$ to be negative for some frequencies, so that $S(\nu,\theta)$ dips below the quantum limit at those frequencies, even though there is no two-mode squeezing in the total field. Moreover, we see that entanglement occurs for all frequencies and the maxima of entanglement correspond to the minima of the variances $S(\nu,\theta)$ and $S(\nu,\theta+\pi/2)$. The maxima of entanglement occur at frequencies corresponding to the imaginary parts of the roots of the $D_{j}(\nu)$ polynomials.  Note that the entanglement that lies in the range of low frequencies $(|\nu|\leq 1)$ is attributed to squeezing in the $\theta$ quadrature component of the output field, $S(\nu,\theta)$, whereas the entanglement that lies in the range of high frequencies $(|\nu| >1)$ is attributed to squeezing in the $\theta +\pi/2$ component $S(\nu,\theta+\pi/2)$.
\begin{figure}[hbt]
\begin{center}
\begin{tabular}{c}
\includegraphics[height=12cm,width=0.9\columnwidth]{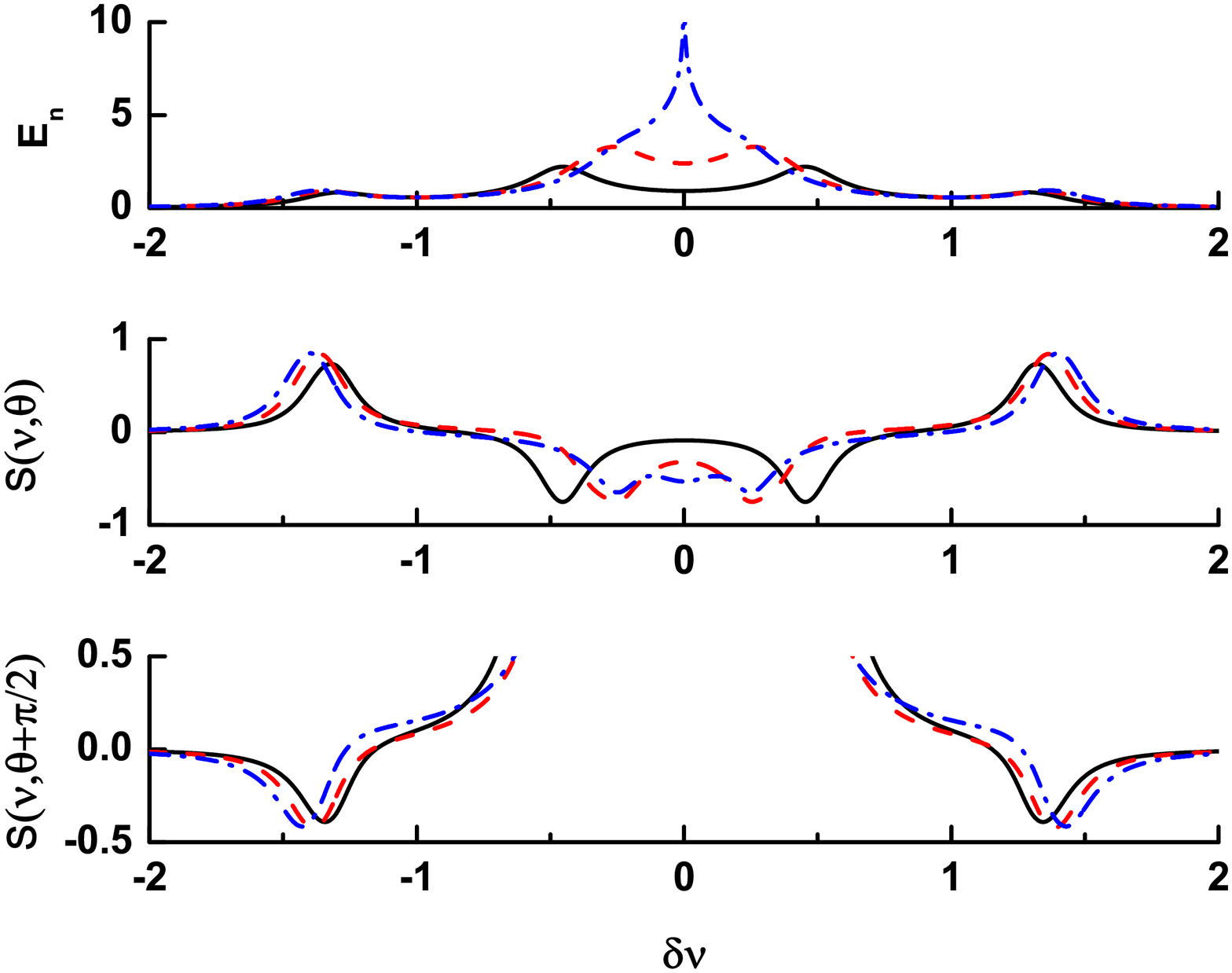}
\end{tabular}
\end{center}
\caption[nsa] {\label{fig7} The variation of the logarithmic negativity $E_{n}(\nu)$ and the variances $S(\nu,\theta)$ with the frequency interval $\delta\nu =\bar{\omega}_{l}-\nu$ for fixed $\omega_0=\omega=1,\delta=0.1 \pi,\alpha_{k}=0.1$, $\kappa=0.2$ and several different values of $\beta$: $\beta=0.4,\theta=1.6958$ (solid line), $\beta=0.46,\theta=1.6734$ (dashed line), $\beta=0.4932,\theta=1.7625$ (dotted line).}
\end{figure}

It is interesting to observe that the cavity modes can be entangled regardless of the size of the atomic ensemble. However, the frequency range at which the modes are entangled varies with the finite-size parameter~$\alpha_{k}$, i.e., the modes can be entangled at several different frequencies. When $\alpha_{k}=0$, the logarithmic negativity and the variances can both have four peaks. When $\alpha_{k}\neq 0$, the peaks merge towards $\delta\nu =0$. In this case, two-mode squeezing occurs only in the $\theta$ quadrature. It follows that with a finite-size ensemble, the largest entanglement is observable in principle at the frequency~$\bar{\omega}_{l}$.  Thus, the size effect is to concentrate the entanglement at the central component of the spectrum. In other words, an optimal degree of squeezing (entanglement) at $\delta\nu =0$ indicates that the output modes are correlated about the average laser frequency $\bar{\omega}_{l}$.

Figure~\ref{fig7} illustrates how the pumping strength $\beta$ alters spectral redistribution of two-mode squeezing and entanglement. Again, each curve is plotted with $\theta$ optimized to give best squeezing. It can be seen that only the low frequency entanglement and two-mode squeezing $(|\nu|\leq 1)$ shifts towards the central frequency as~$\beta$ increases. Notice significant variations of the low frequency entanglement and squeezing with relatively small variations of~$\beta$. In contrast, the high frequency entanglement and squeezing are almost insensitive to $\beta$. The effect of raising~$\beta$ causes only a slight shift of the peaks. Thus, the entanglement and two-mode squeezing associated with the $\theta+\pi/2$ quadrature component are less sensitive to $\beta$ than those associated with the $\theta$ quadrature component. Nevertheless, this does not prevent us from achieving the largest entanglement at the cavity frequency.

\section{Conclusions}\label{sec4}

We have shown that finite sizes of atomic ensembles coupled to counter-propagating modes of a high-$Q$ cavity do have a non-negligible effect on coherence, correlations and entanglement between the cavity modes. In particular, we have shown that finite sizes of the atomic ensemble may result in nonorthogonality of the collective bosonic modes. We have found that the mode nonorthogonality can create the first-order coherence between the modes and appears as the transfer mechanism of the fluctuations between the superpositions of the cavity counter-propagating modes. We have shown that the mode nonorthogonality may result in the second-order correlations that are stronger than that of a thermal field. The correlations are manifested in the photon super-bunching effect.
In addition, the nonorthogonality creates correlations between the modes that are necessary for two-mode squeezing and entanglement between the modes. However, we have found that the correlations created are not strong enough to violate the Cauchy-Schwartz inequality and to produce entanglement between the modes. Therefore, we have also considered the spectral distributions of the logarithmic negativity and the variances of the output cavity fields and have found that entanglement, although not present in the total field, can be created between spectral components of the output cavity fields. The effect of the mode nonorthogonality is to concentrate the entanglement at the central component of the spectrum.

\ack
This work was supported by the National Natural Science Foundation of China (Grant Nos. 60878004 and 11074087), the Ministry of Education under project SRFDP (Grant No. 200805110002), and the Natural Science Foundation of Hubei Province (Grant No. 2010CDA075), and the Nature Science Foundation of Wuhan City (Grant Nos. 201150530149).

\appendix
\section*{Appendix A}\label{appa}
\setcounter{section}{1}

In this appendix we shall discuss in detail the derivation of the effective Hamiltonian~(\ref{he2}). The method is based on the adiabatic elimination of the excited atomic states valid in the limit of large detunings of the laser fields and cavity mode frequencies from the atomic transition frequencies (dispersive regime). The method is well known in the literature and more details can be found in Refs~\cite{sch01,gk05,sem08}.

We shall derive an effective Hamiltonian in a dispersive regime of the interaction of the atoms with the cavity modes and external laser fields. Employing the unitary transformation $\hat{U}(t)=\exp[-i(\hat{H}_{0}^{\prime}/\hbar)t]$ to the total Hamiltonian~(\ref{hh1}), where $\hat{H}_{0}^{\prime}$ is given~by
\begin{eqnarray}
\hat{H}_{0}^{\prime} &= \hbar\bar{\omega}_{l}\left(\hat{a}_{R}^\dag \hat{a}_{R} +\hat{a}_{L}^\dag \hat{a}_{L}\right) +\hbar \sum\limits_{j=1}^N\left[\left(\omega_{lu}+\omega_{d}\right)\ket{u_{j }}\bra{u_{j }}\right. \nonumber\\
&+\left. \omega_{ls}|s_{j }\rangle\langle s_{j }| +\omega_{d}|1_{j }\rangle\langle 1_{j }|\right] ,\label{A1a}
\end{eqnarray}
we obtain
\begin{eqnarray}
\hat{H} = \hat{U}^{\dag}(t)\hat{H}_{T}\hat{U}(t) = \tilde{H}_{0} +\tilde{H}_{AL} +\tilde{H}_{AC} ,\label{A1}
\end{eqnarray}
where
\begin{eqnarray}
\tilde{H}_{0} &=&  \hat{U}^{\dag}(t)\hat{H}_{0}\hat{U}(t) = \hbar\Delta_{c} \left(\hat{a}_{R}^\dag \hat{a}_{R} +\hat{a}_{L}^{\dag} \hat{a}_{L}\right) \nonumber\\
&+& \hbar\sum\limits_{j=1}^N \left\{\Delta_{u}|u_{j}\rangle\langle u_{j}|
+\Delta_{s}|s_{j}\rangle\langle s_{j}|
+\left(\omega_{1} -\omega_{d}\right)|1_{j}\rangle\langle 1_{j}|\right\} ,\label{A2}
\end{eqnarray}
\begin{eqnarray}
\tilde{H}_{AL} &=& \hat{U}^{\dag}(t)\hat{H}_{AL}\hat{U}(t)\nonumber\\
&=&  \frac{1}{2}\hbar \sum\limits_{j=1}^N \left\{\Omega_{u}(\vec{r}_{j})|u_{j }\rangle\langle 1_{j }| +\Omega_{s}(\vec{r}_{j})|s_{j}\rangle\langle 0_{j}| +{\rm H.c.}\right\} ,\label{A3}
\end{eqnarray}
and
\begin{eqnarray}
\tilde{H}_{AC} &=&  \hat{U}^{\dag}(t)\hat{H}_{AC}\hat{U}(t) = \hbar g \sum\limits_{j=1}^N\left\{\left(\hat{a}_{R}{\rm e}^{i\vec{k}_{c}\cdot\vec{r}_{j}}
+ \hat{a}_{L}{\rm e}^{-i\vec{k}_{c}\cdot\vec{r}_{j}}\right)|u_{j }\rangle\langle0_{j }|\right. \nonumber\\
&+&\left. \left(\hat{a}_{R}{\rm e}^{i\vec{k}_{c}\cdot\vec{r}_{j}}
+ \hat{a}_{L}{\rm e}^{-i\vec{k}_{c}\cdot\vec{r}_{j}}\right)|s_{j }\rangle\langle1_{j }|+{\rm H.c.}\right\} ,\label{A4}
\end{eqnarray}
in which
\begin{eqnarray}
\Omega_{u}(\vec{r}_{j}) = \Omega_{u} {\rm e}^{i(\vec{k}_{l}\cdot\vec{r}_{j} -\phi_{u})} ,\quad
\Omega_{s}(\vec{r}_{j}) = \Omega_{s} {\rm e}^{i(\vec{k}_{l}\cdot\vec{r}_{j} -\phi_{s})} ,\label{A5}
\end{eqnarray}
are the position dependent Rabi frequencies of the laser fields,
\begin{eqnarray}
\Delta_{u} = \omega_{u} -\bar{\omega}_{l} ,\qquad \Delta_{s} =\omega_{s} -\omega_{ls} ,\qquad \Delta_{c} =\omega_{c} -\bar{\omega}_{l} ,\label{A6}
\end{eqnarray}
are detunings of the atomic transition frequencies and of the cavity frequency from the laser field frequencies, with
\begin{eqnarray}
\bar{\omega}_{l} =\frac{1}{2}(\omega_{ls} +\omega_{lu}) ,\qquad \omega_{d} =\frac{1}{2}(\omega_{ls} -\omega_{lu}) ,\label{A7}
\end{eqnarray}
standing for the average frequency and detuning between the laser frequencies, respectively.

We may extract a part of the Hamiltonian $\tilde{H}_{0}$ that involves the upper states of the atoms
\begin{eqnarray}
\tilde{H}_{0}^{\prime\prime} = \hbar\sum\limits_{j=1}^N \left(\Delta_{u}|u_{j}\rangle\langle u_{j}|
+\Delta_{s}|s_{j}\rangle\langle s_{j}|\right) ,\label{A8}
\end{eqnarray}
and make a unitary transformation of the remaining part of the Hamiltonian $\hat{H}$ to obtain
\begin{eqnarray}
&H_{I}(t) = \exp[i(\tilde{H}_{0}^{\prime\prime}/\hbar)t]\left(\hat{H}-\tilde{H}_{0}^{\prime\prime}\right)\exp[-i(\tilde{H}_{0}^{\prime\prime}/\hbar)t] \nonumber\\
&= \hbar\Delta_{c} \left(\hat{a}_{R}^\dag \hat{a}_{R} +\hat{a}_{L}^{\dag} \hat{a}_{L}\right)
+\hbar\sum\limits_{j=1}^N \left(\omega_{1} -\omega_{d}\right)|1_{j}\rangle\langle 1_{j}| \nonumber\\
&= \hbar \sum\limits_{j=1}^{N}\left\{\left[\frac{1}{2}\Omega_{u}(\vec{r}_{j})|u_{j }\rangle\langle 1_{j }| +g\!\left(\hat{a}_{R}{\rm e}^{i\vec{k}_{c}\cdot\vec{r}_{j}}
+ \hat{a}_{L}{\rm e}^{-i\vec{k}_{c}\cdot\vec{r}_{j}}\right)\!|u_{j }\rangle\langle0_{j }|\right]\!{\rm e}^{i\Delta_{u}t}\right. \nonumber\\
&+\left. \left[\frac{1}{2}\Omega_{s}(\vec{r}_{j})|s_{j}\rangle\langle 0_{j}|
+g\!\left(\hat{a}_{R}{\rm e}^{i\vec{k}_{c}\cdot\vec{r}_{j}}+ \hat{a}_{L}{\rm e}^{-i\vec{k}_{c}\cdot\vec{r}_{j}}\right)\!|s_{j }\rangle\langle1_{j }|\right]\!{\rm e}^{i\Delta_{s}t}\right\} +{\rm H.c.} \label{A9}
\end{eqnarray}
Consider now the time evolution operator for the time dependent Hamiltonian $H_{I}(t)$
\begin{eqnarray}
\hat{U}_{I}(t) = 1 -\frac{i}{\hbar}\int_{0}^{t}dt^{\prime}H_{I}(t^{\prime}) -\frac{1}{\hbar^{2}}\int_{0}^{t}dt^{\prime}H_{I}(t^{\prime})\int_{0}^{t^{\prime}}dt^{\prime\prime}H_{I}(t^{\prime\prime}) +\ldots \label{A10}
\end{eqnarray}
The first-order contribution is of the form
\begin{eqnarray}
&&\int_{0}^{t}dt^{\prime}H_{I}(t^{\prime}) = t\left\{\hbar\Delta_{c} \left(\hat{a}_{R}^\dag \hat{a}_{R} +\hat{a}_{L}^{\dag} \hat{a}_{L}\right) +\hbar\sum\limits_{j=1}^N \left(\omega_{1} -\omega_{d}\right)|1_{j}\rangle\langle 1_{j}|\right\} \nonumber\\
&-& i\hbar \sum\limits_{j=1}^{N}\left\{\left[\frac{1}{2}\Omega_{u}(\vec{r}_{j})|u_{j }\rangle\langle 1_{j }| +g\!\left(\hat{a}_{R}{\rm e}^{i\vec{k}_{c}\cdot\vec{r}_{j}}
+ \hat{a}_{L}{\rm e}^{-i\vec{k}_{c}\cdot\vec{r}_{j}}\right)\!|u_{j }\rangle\langle0_{j }|\right]\!\frac{{\rm e}^{i\Delta_{u}t}}{\Delta_{u}}\right. \nonumber\\
&+&\left. \left[\frac{1}{2}\Omega_{s}(\vec{r}_{j})|s_{j}\rangle\langle 0_{j}|
+g\!\left(\hat{a}_{R}{\rm e}^{i\vec{k}_{c}\cdot\vec{r}_{j}}+ \hat{a}_{L}{\rm e}^{-i\vec{k}_{c}\cdot\vec{r}_{j}}\right)\!|s_{j }\rangle\langle1_{j }|\right]\!\frac{{\rm e}^{i\Delta_{s}t}}{\Delta_{s}}\right\} +\ldots \label{A11}
\end{eqnarray}
in which the first term is linear in time and the other terms oscillate with detunings $\Delta_{u}$ and $\Delta_{s}$, respectively.

The second-order contribution can be written as
\begin{eqnarray}
&&\int_{0}^{t}dt^{\prime}H_{I}(t^{\prime})\int_{0}^{t^{\prime}}dt^{\prime\prime}H_{I}(t^{\prime\prime}) \nonumber\\
&=& t\hbar^{2}\sum\limits_{j=1}^{N}\left\{\left[\frac{1}{2}\frac{g\Omega_{u}^{\ast}(\vec{r}_{j})}{\Delta_{u}}\left(\hat{a}_{R}{\rm e}^{i\vec{k}_{c}\cdot\vec{r}_{j}}
+ \hat{a}_{L}{\rm e}^{-i\vec{k}_{c}\cdot\vec{r}_{j}}\right)\!|1_{j }\rangle\langle0_{j }| +{\rm H.c.}\right]\right. \nonumber\\
&+&\left. \left[\frac{1}{2}\frac{g\Omega_{s}^{\ast}(\vec{r}_{j})}{\Delta_{s}}\left(\hat{a}_{R}{\rm e}^{i\vec{k}_{c}\cdot\vec{r}_{j}}+ \hat{a}_{L}{\rm e}^{-i\vec{k}_{c}\cdot\vec{r}_{j}}\right)\!|0_{j }\rangle\langle1_{j }| +{\rm H.c.}\right]\right. \nonumber\\
&+&\left. \frac{g^{2}}{\Delta_{u}}\left(\hat{a}_{R}^\dag \hat{a}_{R} +\hat{a}_{L}^{\dag} \hat{a}_{L}+\hat{a}_{L}^\dag\hat{a}_{R}{\rm e}^{2i\vec{k}_{c}\cdot\vec{r}_{j}} + \hat{a}_{R}^\dag\hat{a}_{L}{\rm e}^{-2i\vec{k}_{c}\cdot\vec{r}_{j}}\right)|0_{j }\rangle\langle0_{j }|\right. \nonumber\\
&+&\left. \frac{g^{2}}{\Delta_{s}}\left(\hat{a}_{R}^\dag \hat{a}_{R} +\hat{a}_{L}^{\dag} \hat{a}_{L}+\hat{a}_{L}^\dag\hat{a}_{R}{\rm e}^{2i\vec{k}_{c}\cdot\vec{r}_{j}} + \hat{a}_{R}^\dag\hat{a}_{L}{\rm e}^{-2i\vec{k}_{c}\cdot\vec{r}_{j}}\right)|1_{j }\rangle\langle 1_{j }|\right. \nonumber\\
&+&\left. \frac{1}{4}\frac{|\Omega_{u}|^{2}}{\Delta_{u}}|1_{j }\rangle\langle 1_{j }| +\frac{1}{4}\frac{|\Omega_{s}|^{2}}{\Delta_{s}}|0_{j }\rangle\langle 0_{j }|\right\} +\ldots \label{A12}
\end{eqnarray}
where we extracted only terms that are linear in time.

If we assume that the detunings of the laser fields are much greater than the Rabi frequencies, the cavity coupling constants and the atomic spontaneous emission rates
\begin{eqnarray}
|\Delta_{u}|,|\Delta_{s}|  \gg \Omega_{u},\Omega_{s},g,\gamma ,\label{A13}
\end{eqnarray}
where $\gamma$ is the spontaneous emission rate of the excited states of the atoms, the oscillatory terms make a negligible contribution to the time evolution operator and, after discarding them, we obtain $\hat{U}_{I}(t)\approx 1 -it\hat{H}_{e}/\hbar$, where
\begin{eqnarray}
\hat{H}_{e} &= \hbar\omega \left(\hat{a}_{R}^\dag \hat{a}_{R} +\hat{a}_{L}^\dag \hat{a}_{L}\right)
+\hbar\, \alpha_{k}\delta \left(\hat{a}_{R}^{\dag}\hat{a}_{L} + \hat{a}_{L}^{\dag}\hat{a}_{R}\right) \nonumber\\
&+\hbar\omega_{0}\hat{J}_z +\left[\frac{\hbar\beta_u}{\sqrt{N}}\left(\hat{a}_{R}^\dag \hat{J}_{-k}
+\hat{a}_{L}^\dag \hat{J}_{+k}{\rm e}^{-i\phi_N}\right)+{\rm H.c.} \right]\nonumber\\
&+\left[\frac{\hbar\beta_s}{\sqrt{N}} \left(\hat{a}_{R}^\dag \hat{J}_{+k}^\dag {\rm e}^{i\phi_N}
+\hat{a}_{L}^\dag \hat{J}_{-k}^\dag\right)+{\rm H.c.} \right]  ,\label{A14}
\end{eqnarray}
in which
\begin{eqnarray}
\hat{J}_{z} = \frac{1}{2}\sum\limits_{j=1}^{N}\left(\ket{1_{j}}\bra{1_{j}} -|0_{j}\rangle\langle 0_{j}|\right) ,\qquad
\hat{J}_{\pm k} = \sum\limits_{j=1}^{N}\ket{0_{j}}\bra{1_{j}}{\rm e}^{i\left(\vec{k}_{l}\pm\vec{k}_{c} \right)\cdot\vec{r}_{j}} \label{A15}
\end{eqnarray}
are position dependent collective atomic operators,
\begin{eqnarray}
\omega =\Delta_{c}+\frac{Ng^2}{\Delta} \qquad {\rm and} \qquad \omega_0=\omega_1-\omega_{d} +\frac{(\Omega_u^2-\Omega_s^2)}{4\Delta} \label{A16}
\end{eqnarray}
are detunings of the cavity field frequency and of the atomic frequency $\omega_{1}$ from the laser frequencies modified by the intensity-dependent Stark shifts,
\begin{eqnarray}
\beta_{u} = \frac{\sqrt{N}g\Omega_{u}}{2\Delta} ,\qquad \beta_{s}=\frac{\sqrt{N}g\Omega_{s}}{2\Delta} \label{A17}
\end{eqnarray}
are the effective Rabi frequencies which quantify the strength of the coupling of the effective two-level system to the cavity modes due to virtual transitions to the highly detuned $\ket{u_{j}}$ and $\ket{s_{j}}$ states, and
\begin{eqnarray}
\alpha_{k}\delta = |\alpha_{k}| \frac{Ng^{2}}{\Delta} ,\label{A18}
\end{eqnarray}
with
\begin{eqnarray}
\alpha_{k} = |\alpha_{k}| {\rm e}^{\pm i\phi_{N}} = \frac{1}{N}\sum\limits_{j=1}^{N} {\rm e}^{\pm 2i\vec{k}_{c}\cdot \vec{r}_{j}} .\label{A19}
\end{eqnarray}
In the derivation of Eq.~(\ref{A14}), we have chosen $\Delta_{u}=\Delta_{s}\equiv \Delta$, which involves no loss of generality, and have redefined the cavity field operators that now read $\hat{a}_{R}\equiv \hat{a}_{R}\exp(-i\phi_N/2)$ and $\hat{a}_{L}\equiv \hat{a}_{L}\exp(i\phi_N/2)$. We have assumed further that the laser phases $\phi_u=-\phi_s=\phi_N/2$, where the phase $\phi_{N}$ is defined in~(\ref{A19}). It should also be noted here that the assumption of equal detunings of the laser fields and the cavity modes from the atomic upper states, as illustrated in Fig.~\ref{fig1b}, gives $\Delta_{c}=\omega_{1}-\omega_{d}$. However, due to the presence of the Stark shifts, this condition is  modified to $\omega =\omega_{0}$.

\appendix
\section*{Appendix B}\label{app}
\setcounter{section}{2}

Here, we present the steady-state solutions for the cavity and the collective bosonic modes occupation numbers, average amplitudes and correlations. We assume that all modes were initially in the vacuum state.
In this appendix we present the steady-state solutions for the cavity and the collective bosonic modes occupation numbers, average amplitudes and correlations. We assume that all modes were initially in the vacuum state.
\begin{eqnarray}
\langle \hat{a}^{\dag}_j \hat{a}_j\rangle &=& \frac{\lambda^{2}_{j} (\kappa^2+\Omega^2_j)}{2 \Omega_j h_{j}} ,\nonumber\\
\langle \hat{d}^{\dag}_j \hat{d}_j\rangle &=& \frac{\left\{2 \lambda^2_j \Omega_j+\omega_0 \left[\kappa^{2} + (\omega_0-\Omega_j)^2\right]\right\} h_{j}+8 \lambda^4_1 \Omega^2_j}{4\omega^2_0 \Omega_j h_{j}} ,\nonumber\\
\langle \hat{d}^{\dag}_j \hat{a}_j\rangle &=& -\frac{\lambda_{j} \left[(\Omega_j+i\kappa)(\kappa^2+\Omega^2_j)-h_{j}\right]}{4\Omega_j h_{j}} ,\nonumber\\
\langle \hat{d}_j \hat{a}_j\rangle &=& (-1)^{j}\frac{\lambda_{j} \left[\left(\Omega_j+i\kappa\right)\left(\kappa^2+\Omega^2_j\right)+h_{j}\right]}{4\Omega_j h_{j}} ,\nonumber\\
\langle \hat{a}^2_{j}\rangle &=& (-1)^{j+1}\, \frac{\lambda^2_j (\Omega_j+i\kappa)^2}{2\Omega_j h_{j}} ,\nonumber\\
\langle\hat{d}^2_{j}\rangle &=& \frac{\lambda^2_j (\kappa^2+\Omega^2_j)}{2\omega_0 h_{j}},\label{B1}
\end{eqnarray}
with
\begin{eqnarray}
h_{j} = \omega_0 (\kappa^2+\Omega^2_j) - 4 \lambda^2_j \Omega_j ,\qquad j=1,2.
\end{eqnarray}


\section*{References}
\begin{thebibliography}{99}

\bibitem{kp03} Braunstein S L and Pati A K 2003  {\it Quantum Information with Continuous Variables} (New York: Kluwer)

\bibitem{jz03} Jing J, Zhang J, Yan Y, Zhao F, Xie C and Peng K 2003 {\it Phys. Rev. Lett.} {\bf 90} 167903

\bibitem{at03} Aoki T, Takei N, Yonezawa H, Wakui K, Hiraoka T, Furusawa A and van Loock P 2003 {\it Phys. Rev. Lett.}  {\bf 91} 080404

\bibitem{pf04} Pfister O, Feng S, Jennings G, Pooser R and Xie D 2004 {\it Phys. Rev. A}  {\bf 70} 020302(R)

\bibitem{bl05} Braunstein S L and van Loock P 2005 {\it Rev. Mod. Phys.} {\bf 77} 513

\bibitem{fc05} Felinto D, Chou C W, de Riedmatten H, Polyakov S V and Kimble H J 2005 {\it Phys. Rev. A}  {\bf 72} 053809

\bibitem{wb10} Willis R T, Becerra F E, Orozco L A and  Rolston S L 2010  {\it Phys. Rev. A} {\bf 82} 053842

\bibitem{mp08} Morrison S and Parkins A S 2008 {\it J. Phys. B: At. Mol. Opt. Phys.} {\bf 41} 195502

\bibitem{hp} Holstein T and Primakoff H 1940  {\it Phys. Rev.} {\bf 58} 1098

\bibitem{de07} Dimer F, Estienne B, Parkins A S and Carmichael H J  2007  {\it Phys. Rev. A}  {\bf 75} 013804

\bibitem{ac10} Alcalde M A, Cardenas A H, Svaiter N F and Bezerra V B 2010 {\it Phys. Rev. A} {\bf 81} 032335

\bibitem{sc10} Svidzinsky A A, Chang J T and Scully M O 2010  {\it Phys. Rev. A} {\bf 81} 053821

\bibitem{krb03} Kruse D, Ruder M, Benhelm J, von Cube C, Zimmermann C,  Courteille P W, Els\"asser T H, Nagorny B and Hemmerich A 2003 {\it Phys. Rev. A} {\bf 67} 051802(R)

\bibitem{na03} Nagorny B, Els\"asser T H, Richter H, Hemmerich A, Kruse D, Zimmermann C and Courteille P H 2003 {\it Phys. Rev. A} {\bf 67} 031401(R)

\bibitem{kl06} Klinner J, Lindholdt M, Nagorny B and Hemmerich A 2006 {\it Phys. Rev. Lett.} {\bf 96} 023002

\bibitem{gr00} Gangl M and Ritsch H 2000 {\it Phys. Rev. A} {\bf 61} 043405\\
 Wu Y, Payne M G, Hagley E W and  Deng L 2004 {\it Phys. Rev. A}  {\bf 69} 063803

\bibitem{ps06} Parkins A S, Solano E and Cirac J I 2006 {\it Phys. Rev. Lett.} {\bf 96} 053602

\bibitem{gr06} Guzman R, Retamal J C, Solano E and Zagury N 2006 {\it Phys. Rev. Lett.} {\bf 96} 010502

\bibitem{cb09} Cola M M, Bigerni D, and Piovella N 2009 {\it Phys. Rev. A} {\bf 79} 053622

\bibitem{li06} Li G X, Wu S P and Huang G M 2005 {\it Phys. Rev. A} {\bf 71} 063817\\
\noindent{Li G X 2006 {\it Phys. Rev. A} {\bf 74} 055801}

\bibitem{lk09} Li G X, Ke S S and Ficek Z 2009 {\it Phys. Rev. A} {\bf 79} 033827

\bibitem{gx10} Li G X and Ficek Z 2010 {\it Optics Commun.} \textbf{283} 814

\bibitem{km10} Krauter H, Muschik C A, Jensen K, Wasilewski W, Petersen J M, Cirac J I and Polzik E S  2010  arXiv:1007.2209v1

\bibitem{es08} Esteve J, Gross C, Weller A, Giovanazzi S and Oberthaler M K 2008 {\it Nature (London)} {\bf 455} 1216

\bibitem{gr10} Gross C, Zibold T, Nicklas E,  Esteve J and Oberthaler M K 2010 {\it Nature (London)} {\bf 464} 1165

\bibitem{re10}  Riedel M F, B\"ohi P, Li Y, H\"onsch T W, Sinatra A and Treutlein P  2010 {\it Nature (London)} {\bf 464} 1170

\bibitem{ls10} Leroux I D, Schleier-Smith M H and  Vuletic V 2010 {\it Phys. Rev. Lett.} {\bf 104} 073602

\bibitem{eb70} Rehler N E and Eberly J H 1971 {\it Phys. Rev. A} {\bf 3} 1735

\bibitem{vs10} Vasilyev D V, Sokolov I V and Polzik E S 2010 {\it Phys. Rev. A} {\bf 81} 020302(R)

\bibitem{ls09} Leandro J F and Semiao F L 2009 {\it Optics Commun.} {\bf 282} 4736

\bibitem{lg11} Lazarou C, Garraway B M, Piilo J and Maniscalco S 2011 {\it J. Phys. B: At. Mol. Opt. Phys.} {\bf 44} 065505

\bibitem{ub10} Uys H, Biercuk M J, VanDevender A P, Ospelkaus C, Meiser D, Ozeri R and Bollinger J J 2010 {\it Phys. Rev. Lett.} {\bf 105} 200401

\bibitem{pc08} Porras D and  Cirac J I 2008  {\it Phys. Rev. A} {\bf 78} 053816

\bibitem{mw95} Mandel L and Wolf E 1995  {\it Optical Coherence and Quantum Optics} (Cambridge: Cambridge University Press)

\bibitem{sv05} Shchukin E and Vogel W 2005 {\it Phys. Rev. Lett.} {\bf 95} 230502

\bibitem{hz07} Hillery M and Zubairy M S 2006 {\it Phys. Rev. Lett.} {\bf 96} 050503

\bibitem{gz00} Gardiner C W and  Zoller P 2000 {\it Quantum Noise} (Berlin: Springer)

\bibitem{ks80} Kazantsev A P, Smirnov V S and Sokolov V P 1980 {\it Opt. Commun.}  {\bf 35} 209

\bibitem{lo84} Collett M J, Walls D F and Zoller P 1984 {\it Opt. Commun.} {\bf 52} 145

\bibitem{ag86} Agarwal G S 1986 {\it Phys. Rev. A} {\bf 33} 2472

\bibitem{hr87} Heidmann A and Reynaud S 1987 {\it J. Mod. Opt.} {\bf 34} 923

\bibitem{ft88} Ficek Z and Tana\'s R 1988 {\it Z. Phys. D} {\bf 9} 27

\bibitem{kd77} Kimble H J, Dagenais M and Mandel L 1977 {\it Phys. Rev. Lett.} {\bf 39} 691

\bibitem{dm78} Dagenais M and Mandel L 1978 {\it Phys. Rev. A} {\bf 18} 2217

\bibitem{ch82} Cresser J D, Hager J, Leuchs G, Rateike M S and Walther H 1982 in {\it Dissipative Systems in Quantum Optics} ed. by R. Bonifacio (Berlin: Springer)

\bibitem{dw87} Diedrich F and Walther H 1987 {\it Phys. Rev. Lett.} {\bf 58} 203

\bibitem{fost00} Foster G T, Mielke S L and Orozco L A  2000 {\it Phys. Rev. A} {\bf 61} 053821

\bibitem{pl86} Pegg D T, Loudon R and Knight P L 1986 {\it Phys. Rev. A} {\bf 33} 4085

\bibitem{jy86} Janszky J and Yushin Y 1987 {\it Phys. Rev. A} {\bf 36} 1288

\bibitem{me05} Macovei M, Evers J and Keitel C H  2005 {\it Phys. Rev. A} {\bf 72} 063809

\bibitem{ak07} Akiba K, Kashiwagi K, Yonehara T and Kozuma M 2007 {\it Phys. Rev. A} {\bf 76} 023812

\bibitem{sw07} Stobi\'nska M and  W\'odkiewicz K 2005 {\it Phys. Rev. A} {\bf 71} 032304

\bibitem{zw91} Zou X Y, Wang L J and Mandel L 1991 {\it Opt. Commun.} {\bf 84} 351

\bibitem{df04} Drummond P D and Ficek Z 2004 {\it Quantum Squeezing} (New York: Springer)

\bibitem{lf03} van Loock P and Furusawa A 2003 {\it Phys. Rev. A} {\bf 67} 052315

\bibitem{ls82} Lugiato L A and Strini G 1982 {\it Opt. Commun.} {\bf 41} 67

\bibitem{oh87} Ou Z Y, Hong C K and Mandel L 1987 {\it J. Opt. Soc. Am. B} {\bf 4} 1574

\bibitem{cg84} Collet M J and  Gardiner C W 1984 {\it Phys. Rev. A} {\bf 30} 1386

\bibitem{si00} Simon R 2000 {\it Phys. Rev. Lett.} {\bf 84} 2726

\bibitem{as04} Adesso G, Serafini A and Illuminati F 2004 {\it Phys. Rev. A} {\bf 70} 022318

\bibitem{gv10} Gr\"unwald P and Vogel W 2010 {\it Phys. Rev. Lett.} {\bf 104} 233602

\bibitem{sch01} Schleich W P 2001 \textit{Quantum Optics in Phase Space} (Weinhein: Wiley)

\bibitem{gk05} Gerry C and Knight P L 2005 \textit{Introductory Quantum Optics} (Cambridge: Cambridge University Press)

\bibitem{sem08} Semiao F L 2008 {\it J. Phys. B: At. Mol. Opt. Phys.} {\bf 41} 081004

\end{thebibliography}
\end{document}